\documentclass[12pt]{article}
\usepackage{amssymb}

\usepackage{times}

\title{Classification of life by the mechanism of genome size evolution}

\author
{Dirson Jian Li\footnote{E-mail: dirson@mail.xjtu.edu.cn.}\ \ \& Shengli Zhang\\
\\
\normalsize{\it Department of Applied Physics, Xi'an Jiaotong
University, Xi'an 710049, China} }

\date{}

\usepackage{graphicx}

\begin{document}

\baselineskip24pt

\maketitle

{\bf \small \begin{center} Abstract \end{center}}

The classification of life should be based upon the fundamental
mechanism in the evolution of life. We found that the global
relationships among species should be circular phylogeny, which is
quite different from the common sense based upon phylogenetic trees.
The genealogical circles can be observed clearly according to the
analysis of protein length distributions of contemporary species.
Thus, we suggest that domains can be defined by distinguished
phylogenetic circles, which are global and stable characteristics of
living systems. The mechanism in genome size evolution has been
clarified; hence main component questions on C-value enigma can be
explained. According to the correlations and quasi-periodicity of
protein length distributions, we can also classify life into three
domains.

\newpage

\section{Background and motivation}

In the absence of any ancient genetic sequences, scientists in the
field of molecular evolution have to figure out reasonable
mechanisms to retrieve the evolutionary history according to the
genetic information of contemporary species. Traditionally, the
basis for a natural taxonomy was provided by complex morphologies
and a detailed fossil record. With the sequencing revolution, we had
a new opportunity to understand the richer and more credible
information on the evolution of life stored in the molecular
sequences. Consequently, the basis for the definition of taxa has
progressively shifted from the organismal to the cellular to the
molecular level. Based upon rRNA sequence comparisons, life on this
planet can be divided into three domains: the Bacteria, the Archaea,
and the Eucarya \cite{3domains}. The differences that separate the
three domains are of a more profound nature than the differences
that separate classical five kingdoms (Monera, Protista, Fungi,
Plantae, Animalia).

The protein length evolution is poorly understood at present. The
protein lengths vary notably both within a proteome and among
species, and the average protein lengths of eukaryotes are longer
than the average protein lengths of prokaryotes in general
\cite{Conservation of protein length1} \cite{Conservation of protein
length2}. But there are factors to increase or to decrease protein
length, and it is still unclear whether in general protein tends to
increase in length. \cite{Book molecular evolution} \cite{Curr opin}
\cite{Akashi}. Abound evidence indicates that there is underlying
order in protein sequence organization. It is generally supposed
that there are various structural and functional units in protein
sequences. Periodicity was observed in protein length distributions
\cite{periodicity} \cite{periodicity2}. There is evidence for
short-range correlation of protein lengths according to
investigation by detrended fluctuation analysis \cite{Morariu2}
\cite{Protein length distributions2}. The correspondence between
biology and linguistics at the level of sequence and lexical
inventories, and of structure and syntax, has fuelled attempts to
describe genome structure by the rules of formal linguistics
\cite{Protein Linguistics} \cite{Searls}. So Zipf's law, originally
found in linguistics, can be used to study the rank-size
distribution of protein lengths \cite{Protein length distributions2}
\cite{Searls}.

We found that protein length distributions can be taken as concise
and comprehensive records of the evolution of life. The protein
length should not be taken as a random quantity. The orders in
protein lengths have been recorded in protein length distributions.
We found profound relationship between the protein length evolution
and genome size evolution. So we may unravel the mechanism of genome
size evolution by the properties of protein length distributions. We
found that the global taxonomy of life can be illustrated as
phylogenetic circles rather than phylogenetic trees. Considering
that phylogenetic circles are stable characteristics of living
systems, we suggest that the circular phylogenetic relationship can
be taken as a new criterion to identify domains.

The motivation of this work is to study the mechanisms in genome
evolution based on properties of protein length distributions;
consequently we can classify life in a global scenario of
phylogenetic circles. We can explain (i) the trend of genome size
evolution at the levels of domains and phyla, (ii) the patterns of
genome size distributions, (iii) the bidirectional driving force in
genome size evolution. At last, we successfully classify life into
three domains based on properties of protein length distributions.

When trying to infer the early history of life according to the
present biological data, we can borrow some smart ideas in physics.
There is an analogy between the study of stellar evolution based on
present experimental data of stellar spectra and the current task to
infer the evolutionary history of life based on the protein length
distributions. Although only the contemporary data can be observed
in both cases, we can take the current states of stars, or of
species, as various stages of their evolution. In the former case in
astronomy, the Hertzsprung-Russell diagram shows a group of stars in
various stages of their evolution according to the relation of
absolute magnitude to stellar color, which is helpful to understand
stellar evolution \cite{HR diagram1} \cite{HR diagram2}. In the
latter case in the study of molecular evolution, some similar
diagrams can also be plotted to show a group of species in various
stages of their evolution based on protein length distributions.

\section{Correlation analysis and spectral analysis of protein length distributions}

{\bf Data collection.} The data process in this paper is based on
the biological data. In most calculations based on biological data
in the paper, the protein length distributions are obtained from the
data of $n=106$ complete proteomes ($n^b=85$ bacteria, $n^a=12$
archaea, $n^e=7$ eukaryotes and $n^v=2$ viruses) in the database
Predictions for Entire Proteomes (PEP) \cite{PEP}. Only in the cases
when we study the detailed properties of genealogical circles and
bifurcation of genome size distribution, the protein length
distributions are obtained from both $n=106$ species in PEP and
$775$ species in the National Center for Biotechnology Information
(NCBI).

We denote $s(\alpha)$ as the genome size of species $\alpha$ and
$\eta(\alpha)$ as the ratio of non-coding DNA to the total genetic
DNA of species $\alpha$. The data of $\eta(\alpha)$ and $s(\alpha)$
are obtained from Ref. \cite{eta}, where there are $54$ species ($6$
eukaryotes, $5$ archaebacteria and $43$ eubacteria) can be also
found in PEP. The gene numbers $N$ are obtained by the numbers of
Open Reading Frames (ORFs) in proteomes in PEP. There are $s(\alpha)
\eta(\alpha)$ base pairs (bp) non-coding DNA and $s(\alpha)
(1-\eta(\alpha))$ bp coding DNA in the genome of species $\alpha$.

{\bf Protein length distribution.} We can definitely obtain the
protein length distribution of a species if the lengths of proteins
in its proteome are known. In calculation of a protein length
distribution, we only concern the protein-coding genes and count
only once for a gene with more than one copies. The transposable
elements contribute little in calculation of protein length
distributions. For instance, there are only dozens of genes appear
to have been derived from transposable elements in human genome
\cite{human_rough}.

The protein length distribution of a species $\alpha$ can be denoted
by a vector
\begin{equation}{\mathbf x(\alpha)}=(x_1(\alpha), x_2(\alpha), ...,
x_k(\alpha), ...),\end{equation} where there are $x_k(\alpha)$
proteins, whose lengths are just $k$ amino acids (a.a.), in the
entire proteome of this species (Fig. 1a). The average protein
length in the proteome of species $\alpha$ can be calculated by the
protein length distribution:
\begin{equation}\bar{l}(\alpha)=\frac{\sum_{k=1}^m k\
x_k(\alpha)}{\sum_{k=1}^m x_k(\alpha)}.\end{equation} And the
standard deviation of protein lengths can be calculated by:
\begin{equation}\Delta l (\alpha)= \sqrt{\frac{\Sigma_{k=1}^m
x_k(\alpha)(k-\bar{l}(\alpha))^2}{\Sigma_{k=1}^m x_k(\alpha)}}
\end{equation} The total protein length distribution of all the
species in PEP is denoted by
\begin{equation}\mathbf X=\sum_{\alpha \in \mbox{\tiny PEP}}
\mathbf x(\alpha).\end{equation} Since there are few quite long
proteins, it is practical to choose a sufficient large protein
length as the cutoff of protein length in the protein length
distributions in the data process. Here, we set the cutoff as
$m=3000$ amino acids (a.a.). Almost all the neglected elements of
protein length distributions, i.e., $x_i(\alpha),\ i>m$, vanish
according to the biological data in PEP, which contribute little in
our data analysis. So our conclusions are free from the choice of
$m$.

A peak in the fluctuations of protein length distribution $\mathbf
x(\alpha)$ can be distinguished when $x_l(\alpha)$ is greater than
both $x_{l-1}(\alpha)$ and $x_{l+1}(\alpha)$. The number of peaks of
protein length distribution $\mathbf x(\alpha)$ can be denoted by
$p(\alpha)$. There is no smoothing for protein length distributions
when counting the number of peaks. So $p(\alpha)$ can be obtained
rigorously for any species whose proteome is know. There is profound
biological meaning for peak number $p$.

{\bf Correlation analysis.} Given any pair of species $\alpha$ and
$\beta$ in PEP, we will find several ways to evaluate the
correlation between the protein length distributions of any pair of
species. Accordingly, we can calculate the corresponding average
correlation between any species and all the species in PEP. The
correlation polar angle $\theta(\alpha)$ of species $\alpha$ is
defined as the angle between vectors $\mathbf x(\alpha)$ and
$\mathbf X$, i.e.
\begin{equation}\theta \equiv \frac{2}{\pi} \arccos(\frac{{\mathbf
x} \cdot {\mathbf X}}{\left|\left|{\mathbf x}\right|\right|
\left|\left|{\mathbf X}\right|\right|}),\end{equation} where the
factor $\frac{2}{\pi}$ is added in order that the value of $\theta$
ranges from $0$ to $1$. The less the value of $\theta(\alpha)$ is,
the closer the average correlation of protein length distribution
for species $\alpha$ is.

The correlation coefficient of protein length distributions between
species $\alpha$ and $\beta$ is defined by \begin{equation}
r(\alpha,\beta)=\frac{\sum_{k=1}^m
(x_k(\alpha)-\bar{x}({\alpha}))(x_k(\beta)-\bar{x}({\beta}))}{\sqrt{\sum_{k=1}^m
(x_k(\alpha)-\bar{x}({\alpha}))^2}\sqrt{\sum_{k=1}^m
(x_k(\beta)-\bar{x}({\beta}))^2}},\end{equation} where
$\bar{x}({\alpha})=\frac{1}{m}\sum_{k=1}^m x_k(\alpha)$. And the
average correlation coefficient of species $\alpha$ can be defined
by \begin{equation}R(\alpha)=\frac{1}{106}\sum_{\beta \in
\mbox{\tiny PEP}} r(\alpha,\beta).\end{equation} The value of
$R(\alpha)$ ranges from $0$ to $1$. The more the value of $R$ is,
the closer the average correlation of protein length distributions
is.

The Minkowski distance between species $\alpha$ and $\beta$ is
defined by
\begin{equation}d(\alpha,\beta)=(\sum_{k=1}^m
\left|\frac{x_k(\alpha)}{\left|\left|\mathbf
x(\alpha)\right|\right|}-\frac{x_k(\beta)}{\left|\left|\mathbf
x(\beta)\right|\right|}\right|^q)^{1/q},\end{equation} where $q$ is
a parameter. And the average Minkowski distance of species $\alpha$
can be defined by
\begin{equation}D(\alpha)=\frac{1}{106}\sum_{\beta \in \mbox{\tiny PEP}}
d(\alpha,\beta).\end{equation} The less the value of $D$ is, the
closer the average correlation of protein length distributions is.

{\bf Spectral analysis.} We can study the order in the fluctuations
in the protein length distributions by the method of spectral
analysis. The discrete fourier transformation of the protein length
distribution $\mathbf x(\alpha)$ is:
\begin{equation} \hat{x}_f(\alpha) = \frac{1}{\sqrt{m}} \sum _{k=1}
^m x_k(\alpha) e^{2 \pi i (k-1)(f-1)/m}.\end{equation} The power
spectrum, i.e., the abstract of the discrete fourier transformation,
is defined as (Fig. 1b)
\begin{equation} {\mathbf y(\alpha)} = \left|\left| \hat{\mathbf
x}(\alpha) \right|\right| = \sqrt{(\mbox{Re}\ \hat{\mathbf
x}(\alpha))^2+(\mbox{Im}\ \hat{\mathbf x}(\alpha))^2}.\end{equation}
The power spectrum $\mathbf y=(y_1,...,y_m)$ is mirror symmetric
between $(y_1,...,y_{m/2})$ and $(y_{m/2+1},...,y_m)$ according to
the properties of discrete Fourier transformation.

The peaks in the power spectrum $\mathbf y(\alpha)$ relate to the
periodicity of fluctuations in the protein length distribution
$\mathbf x(\alpha)$. In the following, we only considered the left
half of the power spectrum $(y_1,...,y_{m/2})$, while the properties
on the right half are alike by mirror symmetry. Besides, we
neglected the power spectrum at very low frequency where the peaks
are always high due to the general bell-shape profiles of protein
length distributions. The high frequency sector refers to the power
spectrum at $f$ near to $m/2$, and the low frequency sector refers
to the power spectrum at $f$ much less than $m/2$. The
characteristic frequency of the highest peak in the left half of the
power spectrum $(y_1(\alpha),...,y_{m/2}(\alpha))$ can be denoted by
$f_c(\alpha)$ (Fig. 1b). Moreover, we can find the top $n_p$ highest
peaks in the fluctuations of the left half of the power spectrum.
The maximum frequency of the frequencies for the above top $n_p$
highest peaks can be denoted by $f_{m}(\alpha)$, whose original
intention is to determine an obvious peak with large frequency. 7a -
7c). And we defined the characteristic period $L_c$ and minimum
period $L_m$ of the fluctuations of protein length distribution as
follows:
\begin{eqnarray} L_c(\alpha)=m/f_c\\
L_m(\alpha)=m/f_m, \end{eqnarray} which are free of the choice of
the cutoff $m$. We chose $n_t=30$ for $f_m'$ and $L_m'$ and $n_t=80$
for $f_m''$ and $L_m''$ in the calculation.

The average power spectra for three domains (Bacteria, Archaea and
Eucarya) are as follows respectively:
\begin{eqnarray}\mathbf y^b=\frac{1}{n^b}\sum_{\alpha \in
\mbox{\tiny Bacteria}} \mathbf y(\alpha)\\
\mathbf y^a=\frac{1}{n^a}\sum_{\alpha \in \mbox{\tiny Archaea}}
\mathbf
y(\alpha)\\
\mathbf y^e=\frac{1}{n^e}\sum_{\alpha \in \mbox{\tiny Eucarya}}
\mathbf y(\alpha).\end{eqnarray}

\section{Calculation of genome size and non-coding DNA content}

{\bf Calculation of genome size.} The genome size evolution is one
of the central problems in the study of molecular evolution because
it is a macroevolutionary question and is helpful to understand the
large-scale patterns in the history of life
\cite{Gregory}\cite{Gregory2}. We had found a close relationship
between genome size $s$ and the correlation $\theta$ of protein
length distributions and non-coding DNA content $\eta$ in a previous
work \cite{Dirson_Cambrian}, hence the genome size of contemporary
species can be calculated by an experimental formula with two
variants. In this paper, we also found a close relationship between
genome size $s$ and the peak number $p$, then we obtained another
single-variant experimental formula to calculate the genome sizes.
According to the relationship between the two formulae, we can
obtain an experimental formula to calculate the non-coding DNA
content $\eta$ only based on the data of coding DNA. This
interesting result infers that the non-coding DNA content depends on
the coding DNA.

In the previous work, we found that the genome size $s$ relates to
two variants: the non-coding DNA content $\eta$ and the correlation
polar angle $\theta$. Hence we had obtained a double-variant
experimental formula to calculate the genome size of a certain
species \cite{Dirson_Cambrian}
\begin{equation}s(\eta,\theta)=s_0
\exp(\frac{\eta}{a}-\frac{\theta}{b}),
\end{equation}
where $s_0=7.96 \times 10^6$ bp, $a=0.165$ and $b=0.176$. The crux
to obtain this formula is to find the proportional relationship
between genome size $s$ and correlation polar angle $\theta$ for
prokaryotes. The biological meaning of $\theta(\alpha)$ is the
average correlation of protein length distributions between species
$\alpha$ and all the other species. Furthermore, we naturally
introduced the second variant $\eta$ in the formula so that this
formula can be generalized for eukaryotes. For the prokaryotes, the
non-coding DNA contents are about $10$ percent, but the correlation
polar angles range from about $0.6$ to $0.1$. For the eukaryotes,
the correlation polar angles are around $0.1$, but the non-coding
DNA contents range from $0.1$ to near $1$. This double-variant
formula is well-predicted not only for prokaryotes but also for
eukaryotes. We also proposed a formula to describe the trend of
genome size evolution \cite{Dirson_Cambrian}
\begin{equation}s(t)=s_0
\exp(\frac{\eta(t)}{a}-\frac{\theta(t)}{b}).
\end{equation}
Thus the dynamic parameter $\eta(t)$ and $\theta(t)$ become
promoting factor and hindering factor in determining the trend of
genome size evolution.

In this paper, we found another single-variant experimental formula
to calculate the genome size. We found that there is an exponential
relationship between genome size $s$ and number of peaks $p$:
\begin{equation}s(p)=s'\exp(\frac{p}{p_0}),\end{equation}
where $s'=8.36\times10^4$ bp and $p_0=70.6$ are determined by least
squares. There is only one variant $p$ in this formula. The
prediction of genome sizes by this formula agrees with the
biological data of genome sizes very well (Fig. 2a). The
single-variant formula is also valid not only for prokaryotes but
also for eukaryotes. Therefore, the genome sizes for both
prokaryotes and eukaryotes can be investigated in a unified
framework.

{\bf Calculate of non-coding DNA content.} In terms of the
relationship between the above two experimental formulae, we
obtained an experimental formula to calculate the non-coding DNA
content:
\begin{equation}\eta(\alpha)=0.938\ \theta(\alpha)+0.00234\
p(\alpha)-0.752,\end{equation} where both $\theta$ and $p$ are
defined only based upon the protein length distributions. The
prediction of the non-coding DNA contents agrees with the
experimental observations (Fig. 2b). According to the formulae to
calculate $s$ and $\eta$, we can calculate the size of coding DNA
$(1-\eta)s$ as well as the size of non-coding DNA $\eta s$ according
to the value of $p$ and $\theta$ for any species. At first thought,
such a result is quite surprising. We can obtain the size of coding
DNA directly from the protein length distribution:
$(1-\eta)s=\Sigma_{i=1}^m i x_i$. There should be no direct evidence
about the size of non-coding DNA according to the protein lengths in
the coding DNA segment.

This result is profound because it shows that there is a close
relationship in sizes between non-coding DNA segment $\eta s$ and
coding DNA segment $(1-\eta)s$. The evolution of non-coding DNA
relates to the evolution of coding DNA, whose functions may relate
closely to the cellular differentiation. The size of non-coding DNA
can not be arbitrary if the protein length distribution $\mathbf x$
is given. The information about coding DNA is stored in the
components $x_i$ of protein length distribution $\mathbf x$, where
the order of components is irrelative to the result of calculation.
The order of these component $x_i$, i.e., the order of fluctuations
of protein length distribution $\mathbf x$ becomes significant for
calculating the size of non-coding DNA. So there is additional
evolutionary information stored in the fluctuations of protein
length distributions.

The variant $p$ can be obtained directly from the protein length
distribution of the species' own, but the other variant $\theta$
depends on the data of protein length distributions of other
species. The crux to define $\theta$ is to calculate the correlation
of fluctuations of protein length distributions $\mathbf x$ and
$\mathbf X$. It indicates that there is a universal mechanism for
the genome size evolution for all the species. So the variant
$\theta$, as an average value of correlation, is essential in
calculation of non-coding DNA content.

{\bf Relationship between $p$ and $\eta$, $\theta$.} Genome size
evolution provides a clear example of hierarchy in action. No
one-dimensional explanation can account for the massive variation in
eukaryotic genome sizes \cite{Gregory2}. The success of the
double-variant formula to calculate the genome size benefits from
the proper choice of two variants $\theta$ and $\eta$. But why can
we also find a formula to calculate genome size with only one
variant $p$? The correlation between genome size $s$ and peak number
$p$ can not be explained trivially by the observation that genome
size and proteome size are correlated. The linear relationship
between $p$ and $\log_{10} s$ shows that $p$ is an intrinsic genomic
property of a species. The relationship between $p$ and $\log_{10}
N$ is non-linear (Fig. 2e). The fluctuations of protein length
distributions can no longer be taken as random fluctuations on the
smooth background, which reflects the complexity of proteome and
relates to the complexity of life.

We can understand the biological meaning of peak number according to
the relationship between the single variant $p$ and the pair of
variants $\eta$ and $\theta$. Firstly, we studied the relationship
between $p$ and $\eta$ based on the biological data (see the
distribution of species in Fig. 2c). We found that there is a
critical value $p_c$ of peak number, which definitely separates
prokaryotes and eukaryotes in the $p-\eta$ plane. For prokaryotes,
$p$ is less than $p_c$ and $\eta$ is about constant. The
distribution of species in the $\eta-p$ plane ($p<p_c$) consists a
rightward triangle, which agrees with another triangle distribution
of prokaryotes in $s-\eta$ plane (Fig. 3f) due to the correlation
between peak number $p$ and genome size $s$. The deviation of $\eta$
from average value $0.1$ becomes smaller and smaller when $p$ goes
to $p_c$. For eukaryotes, $p$ is greater than $p_c$ and $\eta$
increases with $p$. The distribution of species in the $\eta-p$
plane ($p>p_c$) consists a leftward triangle. The deviation of
$\eta$ becomes bigger an bigger when $p$ goes away from $p_c$. So
there are few species in the area $p \sim p_c$. Such a regular
distribution of species in the $p-\eta$ plane indicates a profound
relationship between $p$ and $\eta$. So $p$ can not be meaningless
in biology. Secondly, we studied the relationship between $p$ and
$\theta$ (Fig. 2d). We found that $\theta$ declines with $p$ for
prokaryotes when $p<p_c$, while $\theta$ is about constant for
eukaryotes when $p>p_c$. We point out that $p$ relates closely to
the complexity of life. It was suggested that the non-coding DNA
content $\eta$ indicates the complexity of eukaryotes: the larger
the value of $\eta$ (corresponding to larger $p$) is, the more the
complexity of eukaryotes is \cite{prokaryotic_complexity}. In the
case of prokaryotes, the genome sizes, which are proportional to the
gene numbers, can indicate the complexity of prokaryotes because the
non-coding DNA contents are about the same for prokaryotes. Thus,
the larger genome size (corresponding to larger $p$) is, the more
complexity of prokaryotes is. A natural meaning of peak number $p$
is an index for the complexity of structures of any protein length
distribution. Summarizing the above, we suggest that peak number $p$
indicates the meaning of complexity of life.

In order to understand peak number $p$ more clearly, we deduced its
evolutionary formula according to the formula on the evolutionary
trend of genome size in Ref. \cite{Dirson_Cambrian}:
\begin{equation} s(t)=\left\{ \begin{array}{rr} s_1 \exp t/\tau_1,
& t<T_c \\ s_2 \exp t/\tau_2, & t>T_c \end{array} \right.,
\end{equation}
where $T_c=-560$ Million years (Myr) ($t=0$ for today) and
$\tau_1=644$ Myr, $\tau_2=106$ Myr, $s_1=1.98\times10^7$ bp and
$s_2=1.65\times10^9$ bp. We obtained that there were two stages in
the evolution of peak number:
\begin{equation} p(t)=\left\{ \begin{array}{rrr} \frac{p_0}{\tau_1} t
+ p_0 \ln \frac{s_1}{s'}=0.110 t +386, & t<T_c \\
\frac{p_0}{\tau_2} t + p_0 \ln \frac{s_2}{s'}=0.666 t +698, & t>T_c
\end{array} \right.. \end{equation}
The critical peak number in the evolution is $p_c=p(T_c)=p_0 \ln
\frac{s_0}{s'}=324$. We found that peak number evolves much faster
in the period after $T_c$ than in the period before $T_c$. Since
peak number did not evolve evenly, it can not be regarded as an
independent variant in the evolution. The variant $p$ is underlain
by two variants $\eta$ and $\theta$. So the genome size always needs
two-dimensional explanation.

\section{Phylogenetic circles in  $\Delta l -p$ plane and bifurcation
of genome size distribution in $\Delta l -s$ plane}

{\bf Phylogenetic circles in  $\Delta l -p$ plane.} Previously, we
have explained several main problems in C-value enigma, such as the
genome ranges in taxa and genome size distribution, according to the
two-variant genome size formula \cite{Dirson_Cambrian}
\cite{Gregory}. The biological meaning of this formula can be
understood more clearly when we wrote down its derivative form as
follows
\begin{equation} \frac{\Delta s}{s} = \frac{\Delta \eta}{a} -
\frac{\Delta \theta}{b}.\end{equation} Evidently, there are two
factors in control of the genome size evolution. The first variant
$\eta$ is promoter, whose contribution is measured by $a$, and the
other variant $\theta$ is hinderer, whose contribution is measured
by $b$. The genome size evolution is a bidirectional course, which
may either increase or decrease in the evolution.

We found a miraculous distribution of species in $\Delta l -p$ plane
(Fig. 3a). The eukaryotes, archaea and eubacteria distribute in
three circular areas respectively. The species distribute only on
the edges of the circles, and it is empty within the circles. It is
obvious to form a circle by several samples of eukarya, archaea and
mycoplasma respectively. Even for eubacteria, we can also observe a
distribution with an empty center enclosed by a round boundary. The
two virus are also near to each other. We can conclude that there is
almost no exception of species that disassociate these observed
circles.

The standard deviation of protein length $\Delta l$ and the peak
number in protein length distribution $p$ are pivotal properties in
studying genome size evolution. $\Delta l$ relates to the variation
of protein length by its definition. And we can consider $p \sim
\frac{\eta}{a} - \frac{\theta}{b}$ as the ``net driving force'' in
genome size evolution.

{\bf Global patterns of genomes size variation at the levels of
domains and phyla.} In order to observe the phylogenetic circles in
more detail, we obtained more protein length distributions based on
the biological data of $775$ microbes ($725$ eubacteria, $50$
archaea) in NCBI. The microbial taxonomy in this work is based on
the NCBI taxonomy database \cite{NCBI1} \cite{NCBI2}. Thus, we can
obtain a detailed distribution of microbes in $\Delta l - p$ plane.
At the level of domains, we can also observe two phylogenetic
circles for eubacteria and archaea respectively (Fig. 4a).

Too many proteobacteria (blue legends in Fig. 4a) in the database
disturbed us to discern the phylogenetic circle of eubacteria
easily. So we divide the $725$ eubacteria into two groups: ``the
group of $397$ proteobacteria'' and ``the group of the other $328$
eubateria''. Thus, we can discern the phylogenetic circle of
eubacteria. For the group of ``the other $328$ eubateria'', we can
observe a circular chain composed of $10$ phyla (Firmicutes,
Acidobacteria, Actinobacteria, Cyanobacteria,
Bacteroidetes/Chlorobi, Spirochaetes, Chlamydiae/Verrucomicrobia,
Chloroflexi, Deinococcus-Thermus, Thermotogae) (Fig. 4b). This
circular chain shows that the global picture of the distribution of
eubateria are indeed a phylogenetic circle, although the numbers of
species vary greatly among these $10$ phyla in the database. For the
group of proteobacteria, the species from the five classes
(Alphaproteobacteria, Betaproteobacteria, Gammaproteobacteria,
Deltaproteobacteria and Epsilonproteobacteria) also form an arch of
the phylogenetic circle at the same place of the circular chain of
the other $10$ phyla (Fig. 4c). Especially, the species in the class
of Alphaproteobacteria almost form a closed circular distribution.

According to the detailed observation of phylogenetic circle, we
conjecture that the distribution of species in a same domain form a
closed circle in $\Delta l - s$ plane, while the species in a phylum
or lower taxon only form an arch of the circle. Namely, the global
pattern of genomes size variation at the level of domains can be
described by phylogenetic circles, while the pattern of genomes size
variation at the levels of phyla and lower taxa only reflects local
properties of the corresponding phylogenetic circle.

{\bf Bifurcation of genome size distribution in  $\Delta l -s$
plane.} The distribution of species in $\Delta l- s$ plane is very
interesting, whose shape likes two wings of a butterfly (Fig. 3b).
Considering the close relationship between $p$ and $s$, this
distribution is similar to the one in $\Delta l-p$ plane. We can
obviously observe two asymptotes that depart the plane into four
quadrants. The origin $O$, i.e. the point of intersection of the
asymptotes, corresponds to a special genome size $s^*$ (Fig. 3b).
There is almost no species in the upper and lower quadrants. All
bacteria gather either in left quadrant or right quadrant; archaea
gather also in left or right quadrants, but only in the lower parts;
all eukaryotes gather in the right quadrant, but in the upper part
and far away from $s^*$; and the two virus gather in the left
quadrant, but in the upper part and far away from $s^*$. The
distribution of species in $\Delta l- N$ plane is also similar to
the one in $\Delta l-p$ plane, but the circular shapes become worse
(Fig. 3h).

We also obtained a more detailed distribution in $\Delta l -s$ plane
based on the biological data of the $775$ microbes in NCBI. The
overall shape of the distribution also likes a butterfly (Fig. 4d).
Especially, the distribution of proteobacteria in $\Delta l -s$
plane agrees with a butterfly shape very well (Fig. 4e). We observed
that the distribution of species in groups Alphaproteobacteria,
Betaproteobacteria, Gammaproteobacteria also agree with the
butterfly shape; while the species in groups Deltaproteobacteria and
Epsilonproteobacteria distribute on the right wing and left wing
respectively.

Though the distributions of archaea and eukaryotes obviously deviate
the distribution of eubacteria, the overall distribution of all
species does not violate the butterfly shape. The places of Archaea,
Eubateria and Eukarya indicate that, at the level of domain, the
greater the standard error of the protein length in a proteome is,
the greater the genome size is.

The butterfly shaped distribution of species in $\Delta l -s$ plane
is helpful for us to understand the variation of genome sizes, which
strongly indicates the bidirectionality in genome size evolution.
The genome size corresponding to the connection point of the two
wings in Fig. 4d is approximately the same as the genome size
corresponding to the center of phylogenetic circle in Fig. 4b. In
the evolution of genome size for the closely related species, the
genome size may either increase or decrease. It can be indicated
that the increasing trend is considerably stronger than the
decreasing trend in genome size evolution, because there are
obviously more species on the right wing than on the left wing in
the distribution of species in $\Delta l -s$ plane (Fig. 4d and 4e).

\section{Unravelling the mechanism of genome size evolution}

{\bf Global and local pictures of genome size evolution.} A
distinguished phylogenetic circle can be taken as a natural
definition of a domain. The mechanism for the origin of domains is
quite different from the mechanism for the origin of phyla. There
are two significant events in the evolution of life: the origin of
domains in early stage of evolution and the origin of animal phyla
around Cambrian period \cite{Dirson_Cambrian} \cite{Knoll}. The
phylogenetic circles only exist at the level of domain according to
the distribution of species in $\Delta l - p$ plane. The properties
at the level of domains do not couple with the later evolution at
the levels of phyla and so forth. So phylogenetic circles are stable
characteristics and may exist from the early stage in the evolution
of life to present days. The observation of phylogenetic circles can
be explained by Woese's theory on cellular evolution \cite{Woese2}.
According to his perspective, horizontal gene transfer is the
principal driving force in early cellular evolution. The primitive
cellular evolution is basically communal, and it is not the
individual species that evolve at all. So genome cluster evolution
is more essential than the evolution of an individual genome; and
the study of the origin of a living system is much more valuable
than the study of origin of just one species.

The phylogenetic circles are ``global'' properties, while the
traditional conception of the phylogenetic trees is ``local''.
Although the phylogenetic tree is useful in many circumstances, it
has unfortunately misled us to comprehend the panorama of taxonomy
of life. In the global scenario, we emphasize the phylogenetic
circle should evolve as a whole, while in the local scenario, a
species can evolve freely along continually branching phylogenetic
tree. The underlying mechanism on evolution must be a ``global''
theory. An individual life can never originate and evolve unless it
existed as a phylogenetic circle of primordial life. A new theory of
evolution of life is necessary to understand the evolution of a
cluster of genomes as a whole. Then we may explain the trajectory of
the phylogenetic circles in the $\Delta l - p$ plane in the
evolution. The local properties of a phylogenetic circle is the same
as the properties of phylogenetic tree, but we can not be aware of
the global constraint in the local scenario.

{\bf Dynamics of genome size evolution.} Due to the definite
different topological properties between a circle and a tree, a
mechanism of genome size evolution on a circle will quite differ
from the traditional mechanisms of genome size evolution based on
phylogenetic trees. The genome size can not increase unlimitedly
when evolving in a circular pathway, but it can go to infinity in an
unlimitedly branching pathway. So there is intrinsic mechanism to
reduce the genome size which closely relates to the protein
evolution. According to Eqn. (19) and Eqn. (20), we can calculate
the genome size and the non-coding DNA size by $p$ and $\theta$,
both of which are based on the data of protein length distributions.
These experimental formulae indicate that protein evolution
principally drives the genome size evolution.

The driving force in genome size evolution is bidirectional. Large
scale gene duplications and accumulation of transposable elements
are primary contributors in genome expansion. Whereas the reverse
mechanism to reduce genome size was less understood. According to
our scenario of phylogenetic circles, the global circular
relationship in a domain must constrain the genome expansion. More
explicitly speaking, genome size of a species has to evolve in the
community of the domain; it can not evolve independently. In our
previous work on the trend of genome size evolution, there are two
dynamic factors (promotor $\eta$ and hinderer $\theta$) in
determining the genome size evolution, where $\eta$ corresponds to
the process of polyploidy and accumulation of transportable elements
and $\theta$ indicates the relationships among species. So the
biological meaning of the formula on genome size evolution just
agrees with the explanation based on the scenario of phylogenetic
circles.

{\bf Explanation of genome size distribution.} According to the
bifurcation of genome size distribution in  $\Delta l -s$ plane, it
is easy to explain the patterns of genome size distributions among
taxa. There are two main types of genome size distributions:
single-peak type and double-peak type. For single-peak type, the
species in a certain taxon distribute on only one side of wing of
the butterfly shaped distribution in $\Delta l -s$ plane, so the
outline of the genome size distribution among this taxon has only
one main peak. For double-peak type, the species in a certain taxon
distribute on both wings of the butterfly shaped distribution in
$\Delta l -s$ plane, so the outline of the genome size distribution
among this taxon has two main peaks. For examples, the genome size
distributions for Eukaryotes or for Alphaproteobacteria belong to
double-peak type, and the genome size distributions for Archaea or
for Epsilonproteobacteria belong to single-peak type (Fig. 5). The
genome size distributions for eukaryotic taxa belong to single-peak
type \cite{Gregory} \cite{Leitch}, because eukaryotes all distribute
on the right wing of the butterfly shaped distribution in $\Delta l
-s$ plane.

{\bf On plant genome size evolution.} Recent studies have made
significant advances in understanding the mechanism of plant genome
size evolution, where polyploidy and the accumulation of
transposable element plays significant roles in plant genome
expansion, although less is known about the process for DNA removal
\cite{Leitch} \cite{Bowers} \cite{Lynch} \cite{Wolfe} \cite{Vitte}.
It is reported that ``different land plant groups are characterized
by different C-value profiles, distribution of C-values and
ancestral C-values'' \cite{Leitch}. In the viewpoint of phylogenetic
circles, different land plant groups should situate on the
Eukaryotic phylogenetic circle and form a circular chain that is
similar to the bacterial circular chain in Fig. 4b. So it is natural
to draw the above conclusion, because (i) ancestral C-values should
spread around on the circular chain and (ii) the ``local''
properties on genome size evolution at different places on the
phylogenetic circle should also differ among different plant groups.

The formulae on the trend of genome size evolution can explain the
coding DNA and non-coding DNA interactions in genome evolution.
Entire genome duplication contributes the majority of genome size
increase in plants. Simple chromosome duplication may double the
genome size on the left hand of Eqn. (17) or (19). However, the
values of $\eta$ and $\theta$ or $p$ on the right hand of Eqn. (17)
or (19) keep invariant because the protein length distribution does
not change in simple chromosome duplication. The apparent
contradiction to the trend of genome size evolution will urge the
alternation of coding DNA so that $\eta$ and $p$ tend to increase in
after the chromosome duplication. Such evolutionary pressure agrees
with the experimental observations. After duplication, the two
copies of the gene are redundant. Because one of the copies is freed
from functional constraint, mutations in this gene will be
selectively neutral and will most often turn the gene into a
nonfunctional pseudogene \cite{Gregory}. Hence, the ratio of
non-coding DNA $\eta$ will increase. On the other hand, gene
duplication can provide source of material for the origin of new
genes with to alternative length \cite{Lynch}. Consequently, the
protein length distribution will change to be more complex and the
peak number $p$ will be urged to increase. So $p$ intrinsically
measures the protein evolution. Due to the rapid adjustment shortly
after the chromosome duplication, the genome size can come back to
the trend of genome size evolution as described in Eqns (17) and
(19).

It is also reported that more ancient land plants tended to have
smaller genome sizes \cite{Leitch}. Our theory on genome size
evolution agrees with this experimental observation. According to
Eqn. (21), the overall trend of genome size increased exponentially
with respect to time, so more ancient life tended to have smaller
genome size.

{\bf A roadmap to transform bifurcated distribution in $\Delta l -s$
plane to phylogenetic circles in $\Delta l -p$ plane.} The
distribution of species in $\Delta l -p$ plane is about circular,
while the distribution of species in $\Delta l -s$ plane is about
random in the butterfly shaped area. However, there are intrinsic
relationship between the distributions of species in $\Delta l -p$
plane and in $\Delta l -s$ plane. According to Eqn. (19), we can not
directly explain the deformation form the butterfly shaped
distribution in Fig. 3b to the circular distribution in Fig. 3a. In
the followings, we show that there are interesting relationships
among different schemes of clustering of species based on different
properties such as $\eta$, $\theta$, $s$, $\Delta l$ and $\bar{l}$
etc. The clustering analysis can help us understand the
classification of life.

On one hand, according to the approximately proportional
relationship between $\Delta l$ and $\bar{l}$ (Fig. 3c), it is easy
to understand the similarity between the distribution of species in
$s-\bar{l}$ plane (Fig. 3d) and the distribution in $s-\Delta l$
plane (Fig. 3b). The relationship between $\Delta l$ and $\bar{l}$
can be explained by the fact that all the profiles of the protein
length distributions are similar, which relates to stochastic
process \cite{Morariu1} \cite{stochastic model}. Next, we can
explain the mirror symmetry with respect to a horizontal line
between the distribution in $s-\bar{l}$ plane (Fig. 3d) and the
distribution in $s-\eta$ plane (Fig. 3f) according to the coarse
linear relationship between $\bar{l}$ and $\eta$ for prokaryotes
(Fig. 3e). Furthermore, we know that the distribution of species in
$p-\eta$ plane is similar to the distribution in $s-\eta$ plane
according to Eqn. (19).

In a previous work \cite{Dirson_Cambrian}, we have explained the
transformation from the symmetric distribution of species in
$\eta-\theta$ plane (Fig. 3g) to the asymmetric distribution of
species in $\eta-s$ plane (Fig. 3f) according to Eqn. (17). The
parameters $\eta$ and $\theta$ play promoter and hinderer roles in
genome size evolution. We assume that only parameters in a certain
area in $\eta-\theta$ plane are selected by the mechanism in genome
evolution, which results in the bifurcated distribution in $\Delta l
-s$ plane.

On the other hand, we observed the circular structures in the
distribution of species in $\eta-\Delta l$ plane (Fig. 3i). So, the
distribution of species in $p-\Delta l$ plane becomes circles rather
than random when we transform the distribution of species in
$p-\eta$ plane to the distribution of species in $p-\Delta l$ plane
(Fig. 3a). Thus, we found a chain of transformations from the
distribution of species in $s-\Delta l$ plane to distribution of
species in $p-\Delta l$ plane, which can transform the butterfly
shaped distribution in Fig. 3b to the circular distribution in Fig.
3a.

\section{Classification of life by correlation and quasi-periodicity
of protein length distributions.}

{\bf Cluster analysis of protein length distributions.} We propose a
new method to classify life on this planet, which is based on
cluster analysis of protein length distributions. Unsurprisingly,
our results agree with the proposal of three-domain classification,
because the information in the fluctuations of protein length
distributions also comes from the information in the molecular
sequences. Interestingly, we shown again that the fluctuations of
protein length distributions can not be taken as random
fluctuations, which are essential in clustering species. Some
standard cluster analysis methods in the theory of multivariation
data analysis are applied to classify the protein length
distributions of the species in PEP. We introduced average
correlation efficient $R(\alpha)$, average Minkowski distance
$D(\alpha)$, average protein length $\bar{l}(\alpha)$ and peak
number $p(\alpha)$ etc. for each species $\alpha$ in PEP (see
Definitions and notations). All of the above quantities can be
calculated only based on the data of protein length distributions.
Three domains (Bacteria, Archaea and Eucarya) can be separated
successfully according to the distributions of species in the plots
of the relationships among these quantities.

Firstly, we studied the distribution of species in $\bar{l}-p$
plane, where $\bar{l}$ and $p$ only depend the data of the species's
own. We found that the groups of species in Bacteria, Archaea and
Eucarya cluster together in three regions respectively (Fig. 6a).
The archaea cluster in a small region where $\bar{l}$ and $p$ are
relatively small; the bacteria cluster in a region where $\bar{l}$
and $p$ are relatively middle; and the eukaryotes cluster in a
region where $\bar{l}$ and $p$ are relatively large. Thus, we have a
new method to classify life. If the protein length distribution of a
species is known but its classification if unclear, we can calculate
average protein length $\bar{l}$ and peak number $p$ of this
species. Then we can determine which domain the species belongs to
according to its position in the $\bar{l}-p$ plane. Generally
speaking, there is a correlation for the three domains: large $p$
corresponds to large $\bar{l}$. Such a correlation, however, is
invalid for the species in the same domain. The relationship between
$p$ and the genome size $s$ is much closer than the relationship
between $p$ and average protein length $\bar{l}$ (Comparing Fig. 2a
and Fig. 6a). If considering only one quantity, either $\bar{l}$ or
$p$, we can not separate archaea from bacteria.

Secondly, we studied the distribution of species in $R-\log_{10} D$
plane, where $R$ and $D$ depend the data of other species according
to their definitions, where the groups of species in three domains
also cluster together respectively (Fig. 6b). The cluster of
eukaryotes is separated obviously. The small region of the archaea
borders on the big region of Eubacteria, so Archaea and eubacteria
can still be separated. In the above, we chose the parameter $q=1/4$
in the definition of Minkowski distance $d(\alpha, \beta)$ and
accordingly calculate the average Minkowski distance $D$. According
to this choice of parameter $q$, we can separate the three domains
more easily only by the average Minkowski distance $D$. The results
are alike if varying $q$ from $1/2$ to $1/8$ in calculating $D$.

At last, we studied the distribution of species in other plots.
According to the distributions of species in the plots of
$\bar{l}-R$, $\bar{l}-\log D$ and $D-p$, we found that the groups of
species in three domains still cluster together in the corresponding
plots respectively (Fig. 6c).

{\bf Cross-validated ROC analysis.} The cross-validated receiver
operating characteristic (ROC) analysis can be taken as an objective
measure to check for the quality of the above cluster analysis
\cite{ROC}. For instance, we can check the validity of the method in
the cluster analysis between Bacteria and Archaea by $R$ and
$\log_{10} D$ in the following. We found that the cross-validated
ROC curves deviate from the diagonal line obviously, which shows the
validity of our methods to cluster species according to the
properties of their protein length distributions (Fig. 6d).

The method to draw the cross-validated ROC curve is as follows in
detail. Firstly, we randomly separated the species in PEP into two
groups $G_1$ and $G_2$. There are $42$ bacteria, $7$ archaea, $3$
eukaryotes and $1$ viruses in group $G_1$ and the remaining $53$
species are in group $G_2$ (Fig. 6d). Only based on the biological
data of the species in $G_1$, we can define corresponding average
correlation coefficient $R^*$ and average Minkowski distance $D^*$.
According to the distribution of species in $G_1$ in the
$R^*-\log_{10}D^*$ plane, the boundary between Bacteria and Archaea
can be marked according to the distributions $(R^*,\ \log_{10} D^*)$
for the species in $G_1$. Then we can calculate the correlation
coefficient $r(\alpha, \beta)$ and Minkowski distance $d(\alpha,
\beta)$ between each of the species $\alpha \in G_2$ and the species
$\beta \in G_1$, and accordingly obtain their average values for
each species $\alpha \in G_2$
\begin{eqnarray}R^{**}(\alpha)=\frac{1}{53}\sum_{\beta \in
\mbox{\tiny G}_1}
r(\alpha,\beta)\\D^{**}(\alpha)=\frac{1}{53}\sum_{\beta \in
\mbox{\tiny G}_1} d(\alpha,\beta). \end{eqnarray} Still in the
$R^*-\log_{10}D^*$ plane, we obtained a group of dots $(R^{**},\
\log_{10} D^{**})$ for species in $G_2$. Some of the archaea in
$G_2$ still belong to the region of archaea according to the
boundary defined by the data of species in $G_1$, while other
archaea in $G_2$ cross the boundary. We can obtain the cross
validated ROC curve according to the validity of cluster analysis
for the species in $G_2$ by shifting the position of the boundary
(Fig. 6e). We can repeat the above procedure after changing over the
data between $G_1$ and $G_2$. Then we obtained another
cross-validated ROC curve.

\section{Spectral analysis of protein length distributions}

{\bf Characteristics of power spectrum.} The evolution of protein
length is a virgin field in the study of molecular evolution.
Although the mechanism of the evolution of protein length is
unknown, we observed order in the protein lengths such as the
quasi-periodicity, long range correlation and the tendency for
conservation of protein length in domains. In this paper, we try to
study the properties of protein length distributions by spectral
analysis. In the section of ``Definitions and notations", we defined
a power spectrum $\mathbf y(\alpha)$ for any species $\alpha$. We
defined the characteristic frequency $f_c$ and the maximum frequency
$f_m$, and we also defined the characteristic period $L_c$ and
minimum period $L_m$ of the protein length distribution. For the
domains Bacteria, Archaea and Eucarya, we defined the average power
spectra $\mathbf y^b$, $\mathbf y^a$ and $\mathbf y^e$ respectively.
Considering additional quantities such as average protein length
$\bar{l}$, peak number $p$ and non-coding DNA content $\eta$, we
observed some interesting correlations among these quantities. We
show that there are correlations between protein lengths at
different scales.

{\bf The protein length hierarchy.} Structures can be observed in
the fluctuations of the protein length distributions. We found that
there are correlations between the characteristic frequency $f_c$
and maximum frequency $f_m$ (Fig. 7a, 7c). The characteristic
frequency $f_c$ increases with the maximum frequency $f_m$, which is
especially obvious for archaea and eukaryotes. There is also
correlation between characteristic period $L_c$ and the minimum
period $L_m$ (Fig. 7b, 7d). The values of $L_c$ and $L_m$ are
intrinsic properties of protein length distributions that are free
from the choice of cutoff $m$. Hence we found that the
characteristic period $L_c$ increases with the minimum period $L_m$,
especially for archaea and eukaryotes. Such an intrinsic correlation
between $L_c$ and $L_m$ shows that there is a hierarchy in protein
lengths. There might be a general mechanism in the organization of
protein segments, which results in that the long protein length
period $L_c$ varies with the short protein length period $L_m$ for
individual species.

{\bf The constraint on average protein length.} Comparing the fact
that genome sizes range more than $1,000,000$-fold in the species on
the planet, the average protein lengths in proteomes (several
hundreds a.a.) vary slightly. There is a tendency for conservation
of protein length in Bacteria, Archaea and Eucarya respectively
\cite{Conservation of protein length1} \cite{Conservation of protein
length2}. The average protein lengths in proteomes for Bacteria
range from about $250$ a.a. to about $350$ a.a.; the values for
Archaea are a little smaller; the values for Eucaryotes are around
$500$ a.a.. The protein lengths vary slightly while the genome size
evolves rapidly. Such a sharp contrast awaits answers. One possible
solution is based on the understanding of evolutionary outlines of
genome size and gene number from the beginning of life $t\approx
-3,800$ Myr to present $t=0$. According to the theory in Ref.
\cite{Dirson_Cambrian}, we can obtain a formula of the evolutionary
outline of average protein length
\begin{equation} \bar{l}(t)=\left\{ \begin{array}{llr}
\frac{(1-\eta_1(t))s_1(t)}{3 N_1(t)} & = 242 \exp (-\frac{t}{5320}), & t<T_c \\
\frac{(1-\eta_2(t))s_2(t)}{3 N_2(t)} & = (110-14.3 t) \exp
(\frac{t}{164}), & t>T_c \end{array} \right., \end{equation} where
the subscripts denote two stages in the evolution. This formula can
explain the difference between genome size evolution and protein
length evolution. Genome size increased rapidly, while the average
protein length varied slightly and it even tended to decrease in
each stage of the evolution. Our results agree with experimental
observations in principle. The genome size was approximately
proportional to the gene number before the time $T_c$, so the
average protein lengths for prokaryotes should approximately keep
constant in most time before $T_c$. Then both evolutionary speeds
for gene number and genome size of eukaryotes shifted to new values
after $T_c$, while the coding DNA content $1-\eta$ began to
decrease. Such a transition of evolution of genome size and gene
number around $T_c$ can set an upper limit for the average protein
lengths for eukaryotes in the following evolution. The constraint on
protein lengths could also be explained in an alternative way. The
spectral analysis of protein length distributions might be helpful
for us to understand the intrinsic mechanism of protein length
evolution in detail. According to the relationship between $f_c$ and
$\eta$ and the relationship between $f_c$ and $\bar{l}$, we can
relate the evolution of average protein length $\bar{l}$ to the
non-coding DNA content $\eta$, i.e., $\bar{l}$ tended to decrease
when $\eta$ increased gradually. So the correlation of protein
lengths can intrinsically constrain the average protein lengths in a
certain range in the evolution.

The distribution of species in $f_c-\eta$ plane shows a regular
pattern: the value of $f_c$ tends to go from middle frequency to
either lower frequency or higher frequency when $\eta$ increases
gradually from about $0.1$ to $1$ (Fig. 8b). The same tendency of
$f_c$ can be observed in $f_c-p$ plane (Fig. 8c) when $p$ increases
gradually. The tendency of $f_c$ can be observed clearly especially
according to the distributions of eukaryotes in the above. The
mechanism constraining the average protein length can be inferred by
the rainbow-like distribution of species in $f_c-\bar{l}$ plane,
where the species in Bacteria, Archaea and Eucarya gathered in three
horizontal convex arches respectively (Fig. 8a). Such an order shows
that the average protein length $\bar{l}$ tends to evolve from long
(corresponding middle $f_c$) to short (corresponding lower or higher
$f_c$). We can observe directly that $\bar{l}$ decreases when $\eta$
increases (Fig. 3e), whose intrinsic mechanism, however, should be
revealed by spectral analysis.

{\bf Average power spectra and phylogeny of three domains.} We can
study the properties of average power spectrum for Bacteria, Archaea
and Eucarya respectively, which reflects the phylogeny of three
domains. An important characteristic can be observed that the
bottoms of the profiles of the average power spectra are either
``convex'' or ``concave''. According to the results by several
different ways to smooth the average power spectra $\mathbf y^b$,
$\mathbf y^a$ and $\mathbf y^e$, we always concluded that the
profiles of the average power spectra of Archaea and Eucarya have
``convex bottoms" while the profile of the average power spectrum of
Bacteria has ``concave bottom", where the ``bottom" refers to the
profile of power spectrum at $f$ around $m/2$ (Fig. 9). It is well
known that the relationship between Archaea and Eucarya is closer
than the relationship between Archaea and Bacteria. So the property
of the outlines of the average spectra agrees with the phylogeny of
the three domains. A convex bottom indicates that the power spectrum
in the high frequency sector (at $f\sim 1500$) prevails the power
spectrum in the low frequency sector (at $f\sim 500$); while a
concave bottom indicates the opposite case. So the differences in
the ``bottoms'' of power spectra of three domains might result from
the underlying mechanism of protein length evolution.

In the above, the outlines of the average power spectra are obtained
by smoothing the average power spectra in two methods. In the first
method, we can smoothen $\mathbf y^b$, $\mathbf y^a$ and $\mathbf
y^e$ as followings:
\begin{eqnarray}
\mathbf [Y^b (w)]_f=\frac{1}{2w+1}\sum_{k=f-w}^{f+w}(y^b)_k\\
\mathbf [Y^a (w)]_f=\frac{1}{2w+1}\sum_{k=f-w}^{f+w}(y^a)_k\\
\mathbf [Y^e
(w)]_f=\frac{1}{2w+1}\sum_{k=f-w}^{f+w}(y^e)_k,\end{eqnarray} where
$2w+1$ is the width of the averaging sector and the range of $f$ is
$f=1+w, ..., m-w$. We obtain two sets of outlines of the average
power spectra $\mathbf Y^b(w)$, $\mathbf Y^a(w)$ and $\mathbf
Y^e(w)$ ($w=w_1=100$ or $w=w_2=300$) in the averaging calculations
(Fig. 9a-9c). In the second method, we use the Savitzky-Golay method
\cite{Savitzky-Golay} to obtain outlines of the average power
spectra $\mathbf y_{SG}(\alpha)$ for each species. Then, we averaged
$\mathbf y_{SG}(\alpha)$ for Bacteria, Archaea and Eucarya
respectively and denote the results as $\mathbf Y^b_{SG}$, $\mathbf
Y^a_{SG}$ and $\mathbf Y^e_{SG}$ (Fig. 9d-9f). We found that all the
outlines $\mathbf Y^b(s_1)$, $\mathbf Y^b(s_2)$ and $\mathbf
Y^b_{SG}$ for Bacteria have concave bottoms and the corresponding
outlines for Archaea and Eucarya have convex bottoms.

\section{Conclusion and discussion}

We conclude that the classification of life can be studied according
to the understanding of fundamental mechanism of genome size
evolution. The phylogenetic relationship among species in a domain
is circular rather than the traditional concept of branching trees.
The phylogenetic circle is a global property of living systems at
the level of domain. We propose a natural criterion to define a
domain by each of the phylogenetic circles. We observed at least
three main phylogenetic circles corresponding to three known
domains. In the global scenario of phylogenetic circles, we can
explain the driving force in genome size evolution and the patterns
of genome size distributions. The peak number $p$ plays the role of
net driving force in genome size evolution. The genome size concerns
two factors: (i) the net driving force $p$, and (ii) the circular
phylogenetic relationship in a domain. Thus, there is no trivial
correlation between genome size and biological complexity. The
global circular relationship is quite different from the local
branching relationship. The underlying mechanism in origin and
evolution of life should consider that a domain should evolve as a
whole.

There is rich evolutionary information stored in the fluctuations of
protein length distributions. In the past, the fluctuations in
protein length distribution were routinely assumed as random ones in
a smooth background. Such a prejudice may result in the neglect of
the pivotal evolutionary information stored in the fluctuations of
protein length distributions. Based on the biological data of
protein lengths in a proteome, we can calculate the genome size as
well as the ratios of coding DNA and non-coding DNA for a species.
Our results agree with the biological data very well. So there is
profound relationship between the evolution of non-coding DNA and
the evolution of coding DNA. We reconfirm the three-domain
classification of life by cluster analysis of protein length
distributions. We found that there are correlations between long
periods and short periods of protein length distributions. The
validity of our results can be verified by objective measures, which
shouldn't be ascribed to accidental coincidences. The study on
protein length distributions provides us a chance to understand the
macroevolution of life.

There should be a universal mechanism which underlies the molecular
evolution. The fluctuations in protein length distributions may
result from this universal mechanism. Thus we can determine the
position of a species in the evolution of life by correlation
analysis of the protein length distributions, and therefore obtain a
panorama of evolution of life. There are many analogies between
protein language and natural language of human being. We conjecture
that linguistics may play a central role in the protein length
evolution. A linguistic model was made to study the protein length
evolution. In this model, protein sequences can be generated by
grammars, hence we can obtain simulated protein length distributions
for a set of grammars. The average protein lengths $\bar{l}$ and the
peak numbers $p$ can be calculated consequently. The correlation
between peak numbers $p$ and average protein length $\bar{l}$ in
experimental observation can be explained by the simulation. Our
results indicate an intrinsic relationship between the complexity of
grammars in protein sequences and the peak numbers in protein length
distributions.

\section*{Acknowledgements}

We are grateful to the anonymous reviewers for their valuable
suggestions such as the cross-validated ROC analyses and discussions
on plant genome size evolution. DJL thanks Morariu for discussions.
Supported by NSF of China Grant No. of 10374075.

\clearpage

\begin{figure}
\centering{
\includegraphics[width=70mm]{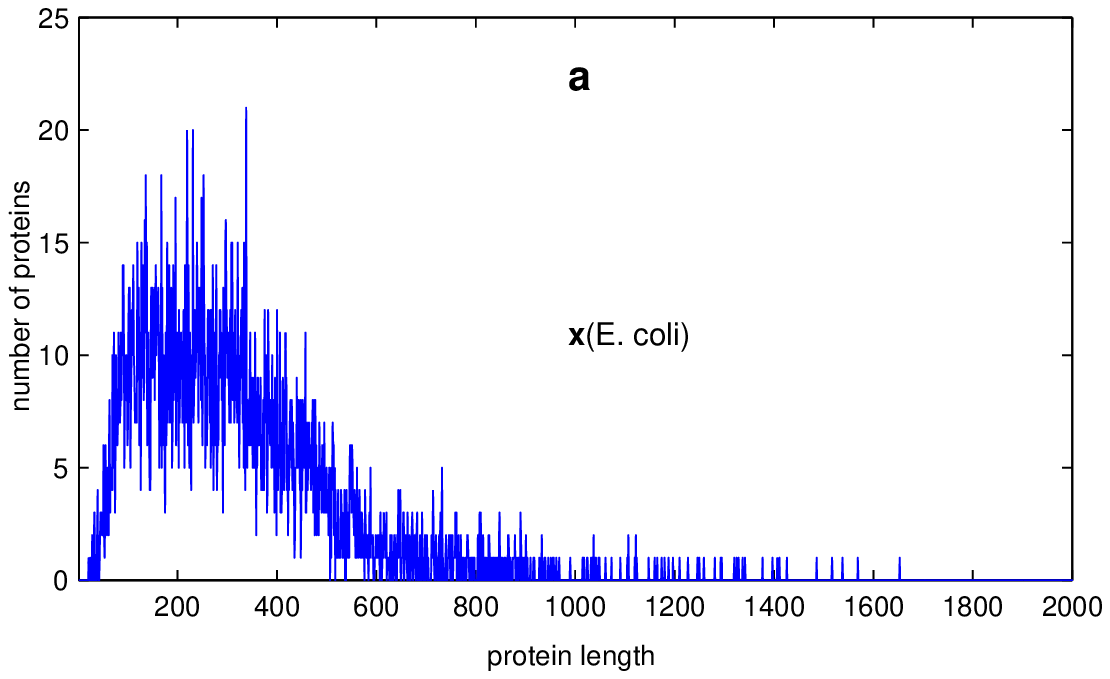}
\includegraphics[width=70mm]{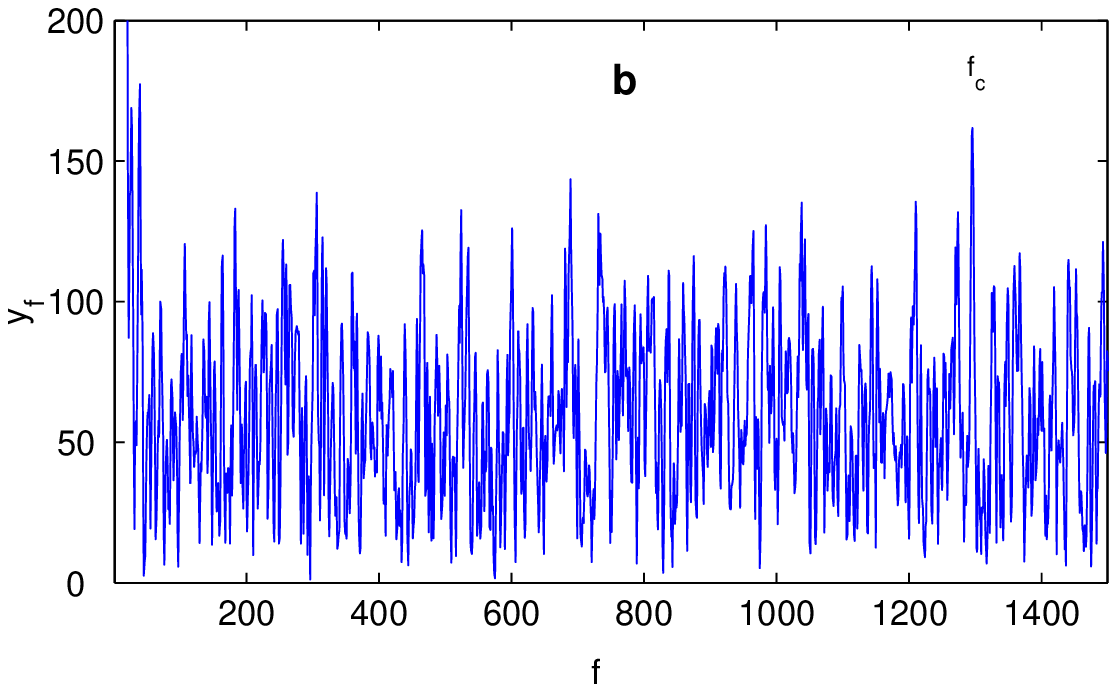}}
\label{fig1} \caption{\small {\bf Protein length distribution and
power spectrum of E. coli.} {\bf a,} Protein length distribution
${\mathbf x}$(E. coli). {\bf b,} The power spectrum ${\mathbf y}$(E.
coli), the characteristic frequency $f_c$ at the highest peak is
marked.}
\end{figure}

\clearpage

\begin{figure}
\centering{
\includegraphics[width=70mm]{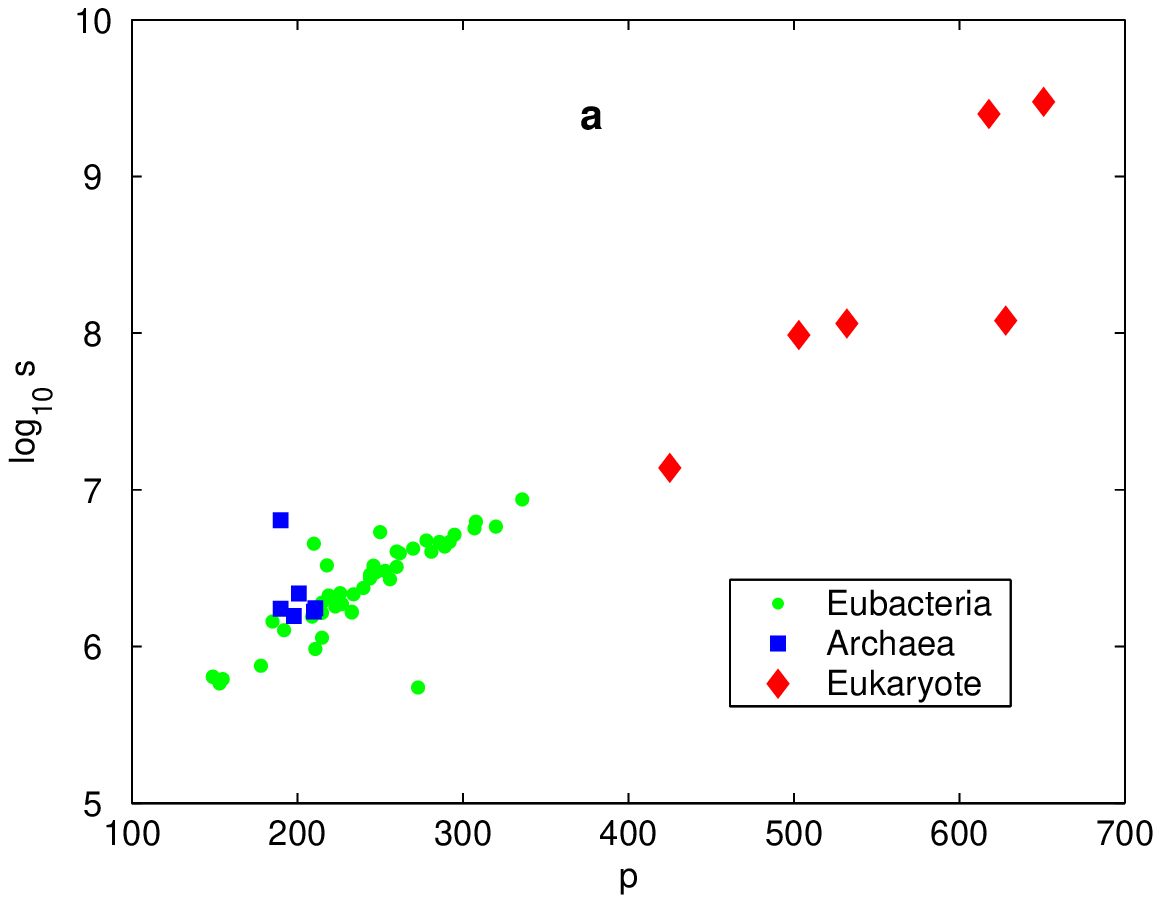}
\includegraphics[width=70mm]{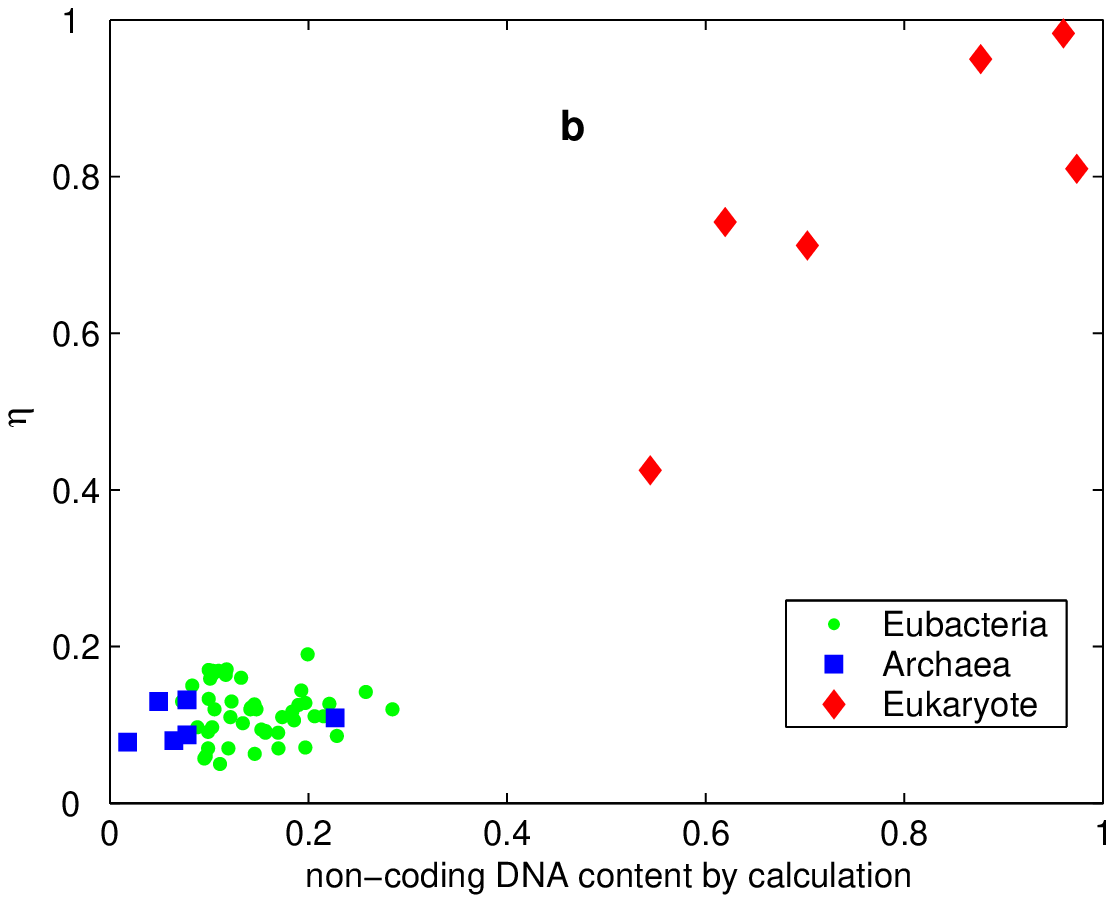}
\includegraphics[width=70mm]{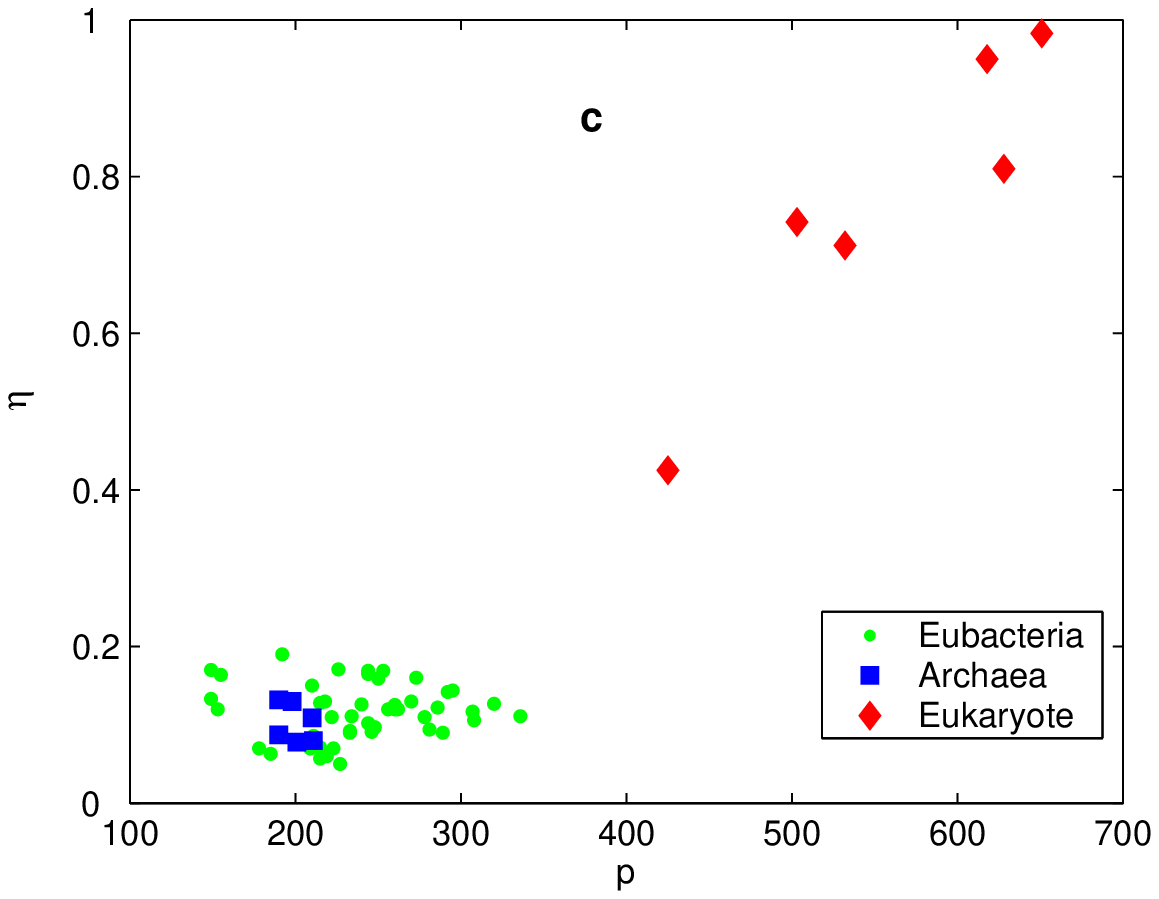}
\includegraphics[width=70mm]{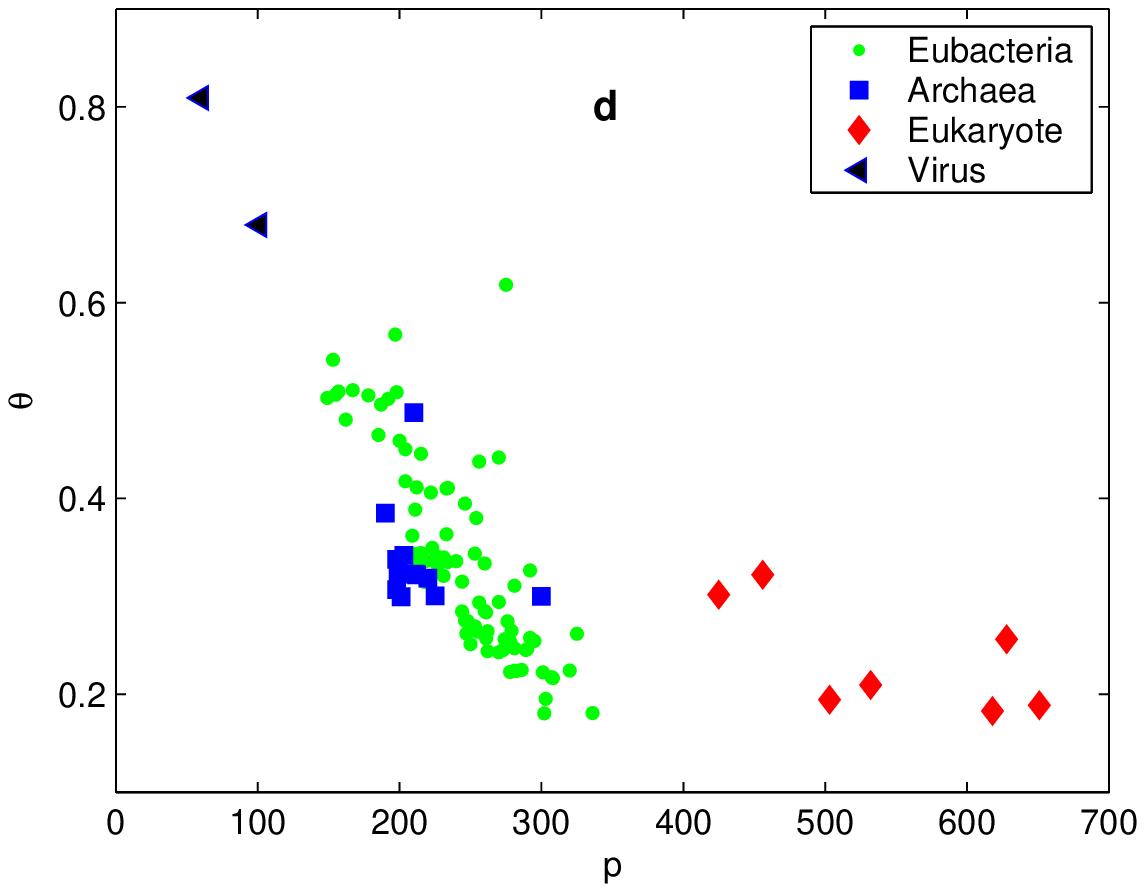}
\includegraphics[width=70mm]{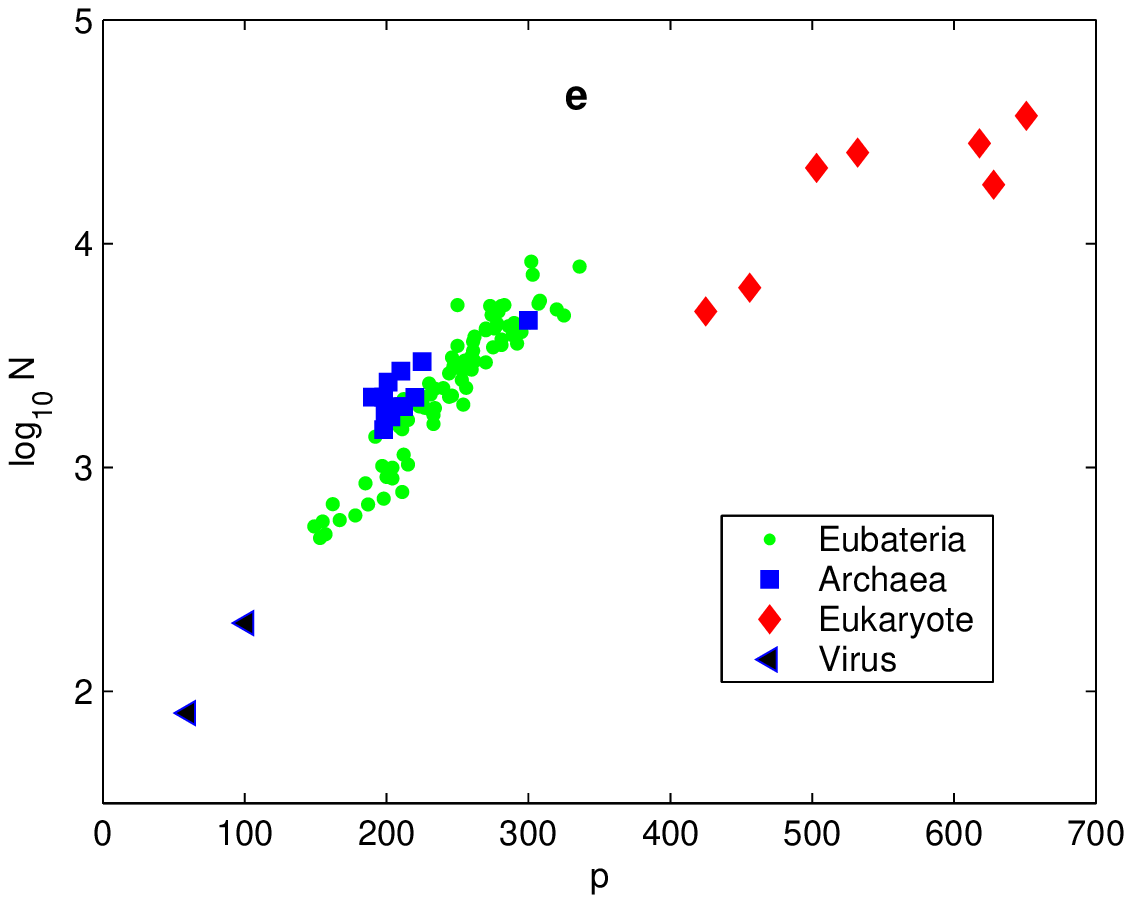}}
\label{fig1} \caption{\small {\bf Prediction of genome size and
non-coding DNA content.} {\bf a,} $p$ is proportional to $\log_{10}
s$ (Correlation coefficient is $0.9428$). {\bf b,} The non-coding
DNA predicted by the formula agrees with the biological data
(Correlation coefficient is $0.9468$). {\bf c,} The relation between
$p$ and $\eta$. {\bf d,} The relation between $p$ and $\theta$. {\bf
e,} The relation between $p$ and $\log_{10} N$ is not linear
(Correlation coefficient is $0.8747$).}
\end{figure}

\clearpage

\begin{figure}
\centering{
\includegraphics[width=160mm]{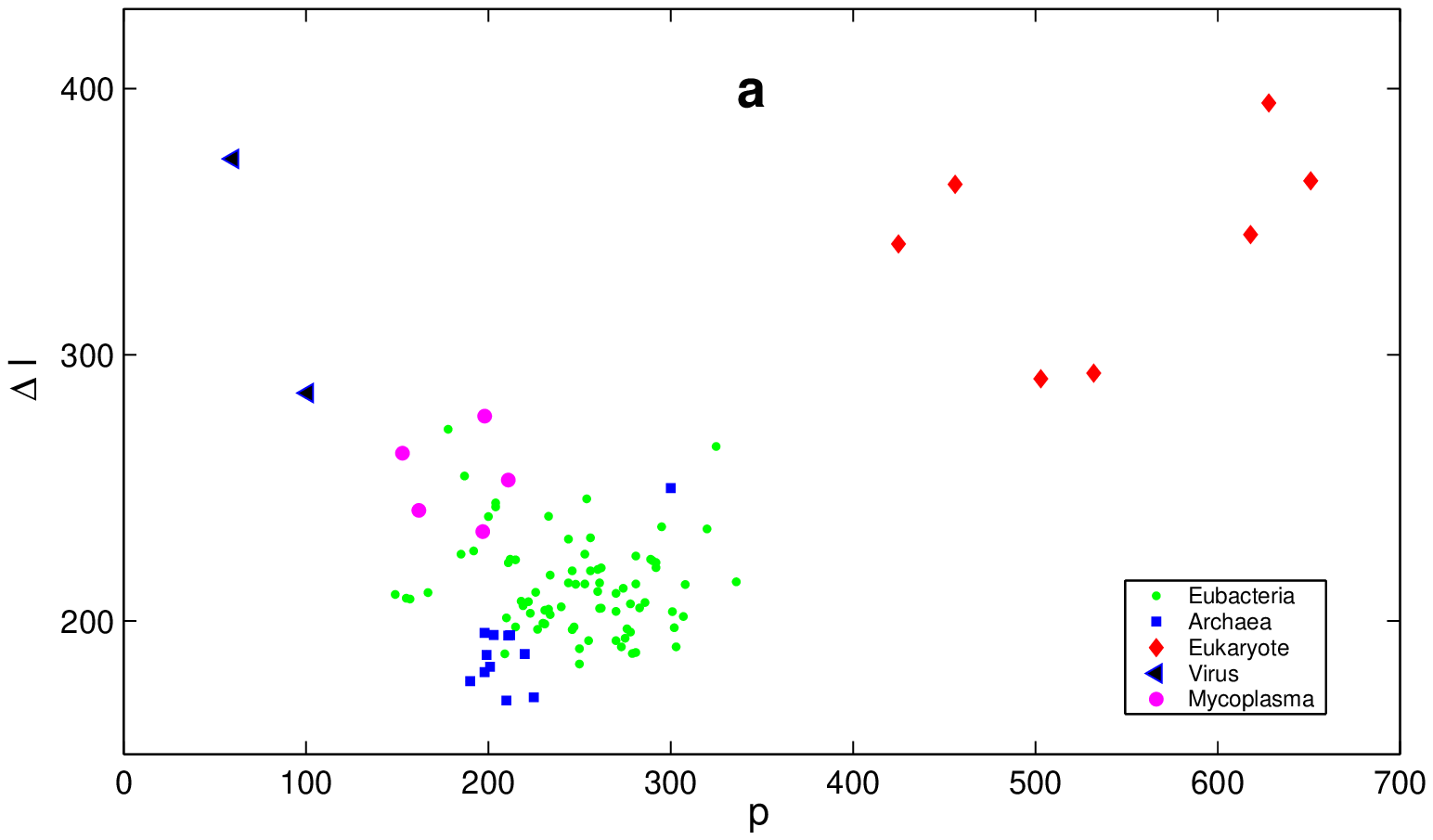}
\includegraphics[width=160mm]{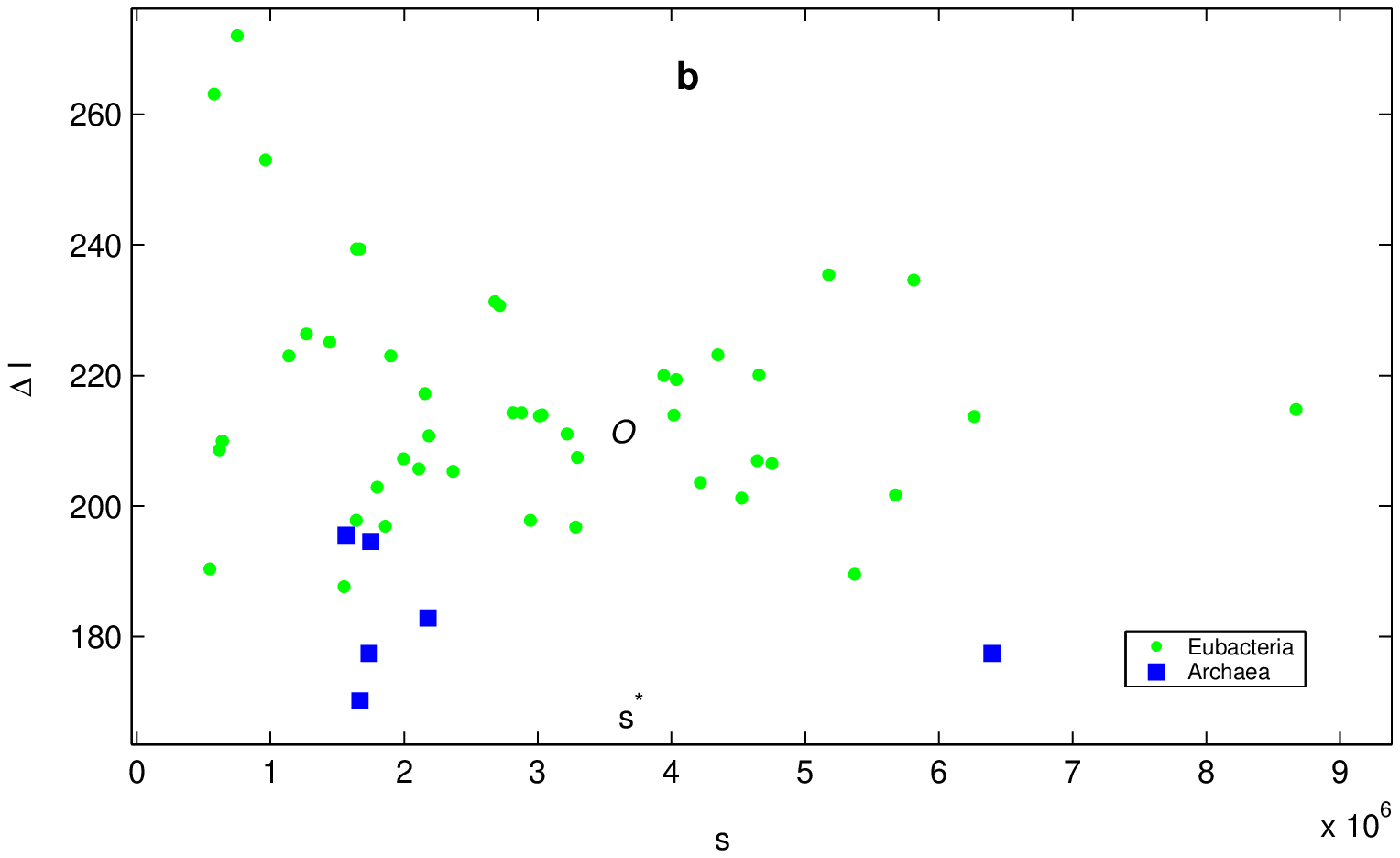}}
\end{figure}
\clearpage
\begin{figure}
\centering{
\includegraphics[width=70mm]{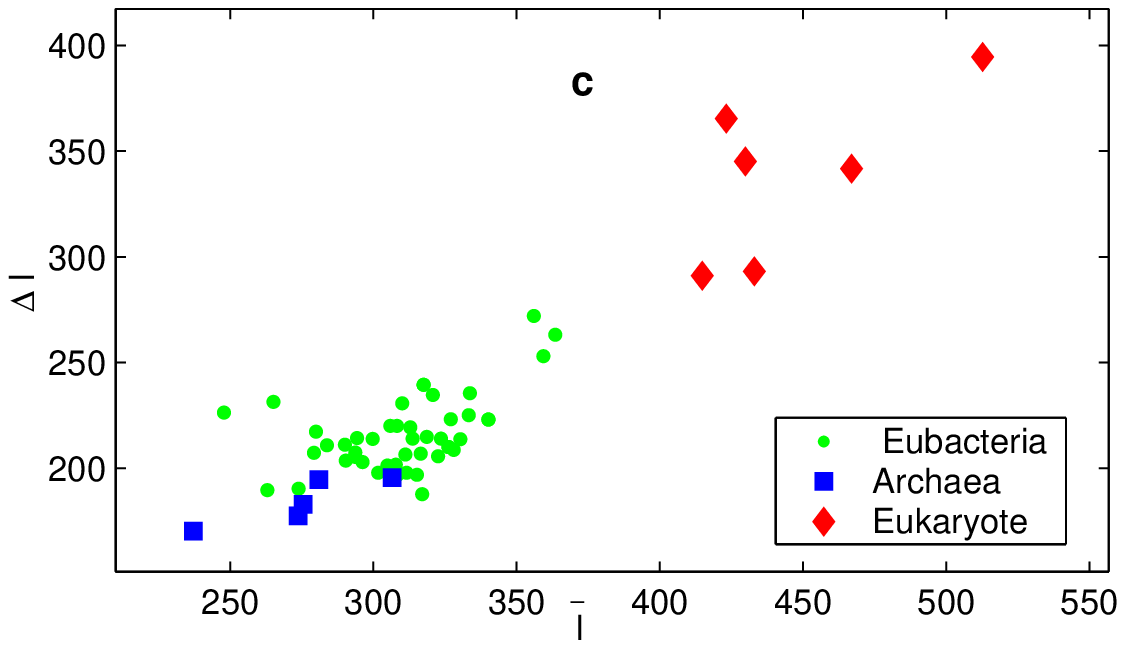}
\includegraphics[width=70mm]{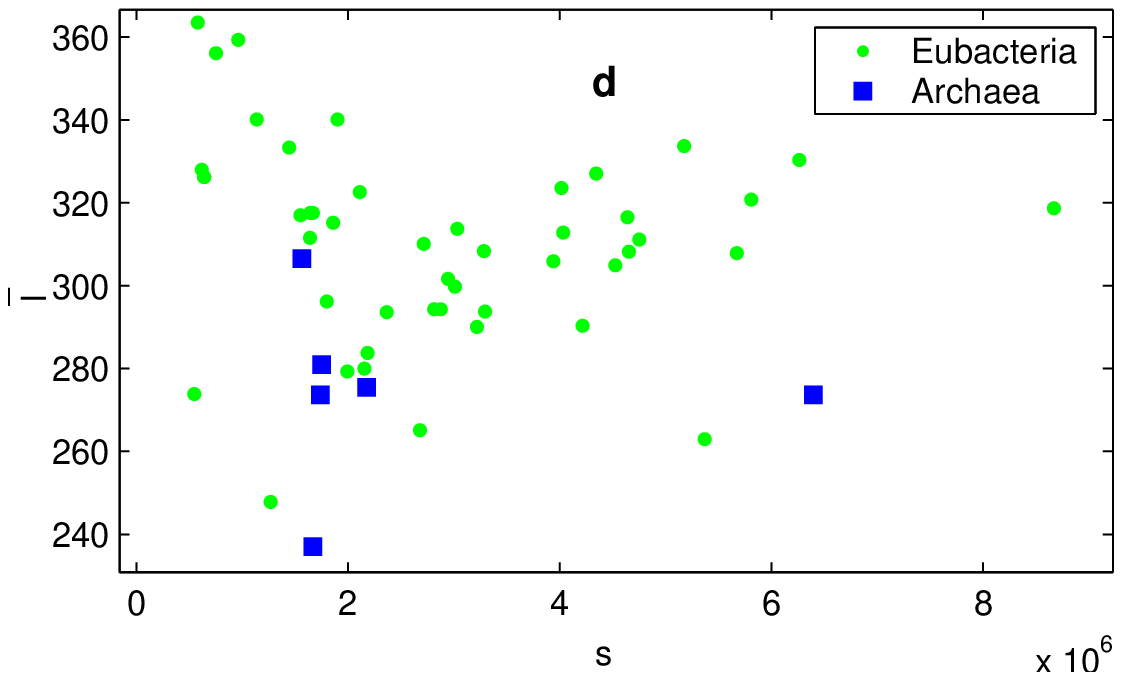}
\includegraphics[width=70mm]{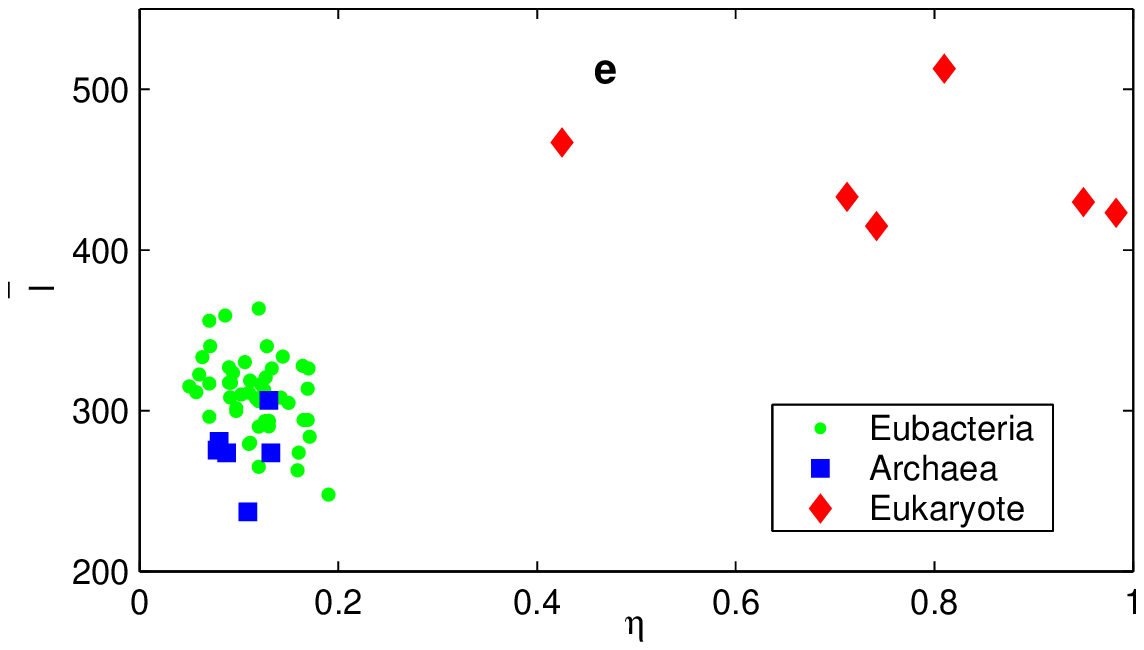}
\includegraphics[width=70mm]{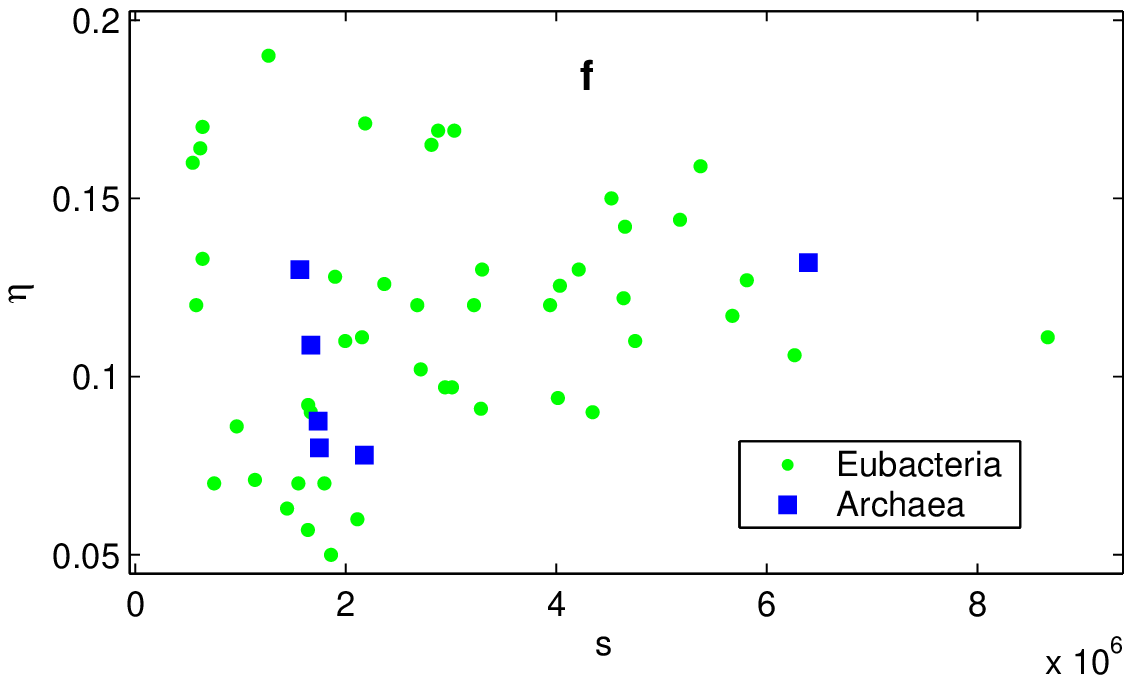}
\includegraphics[width=70mm]{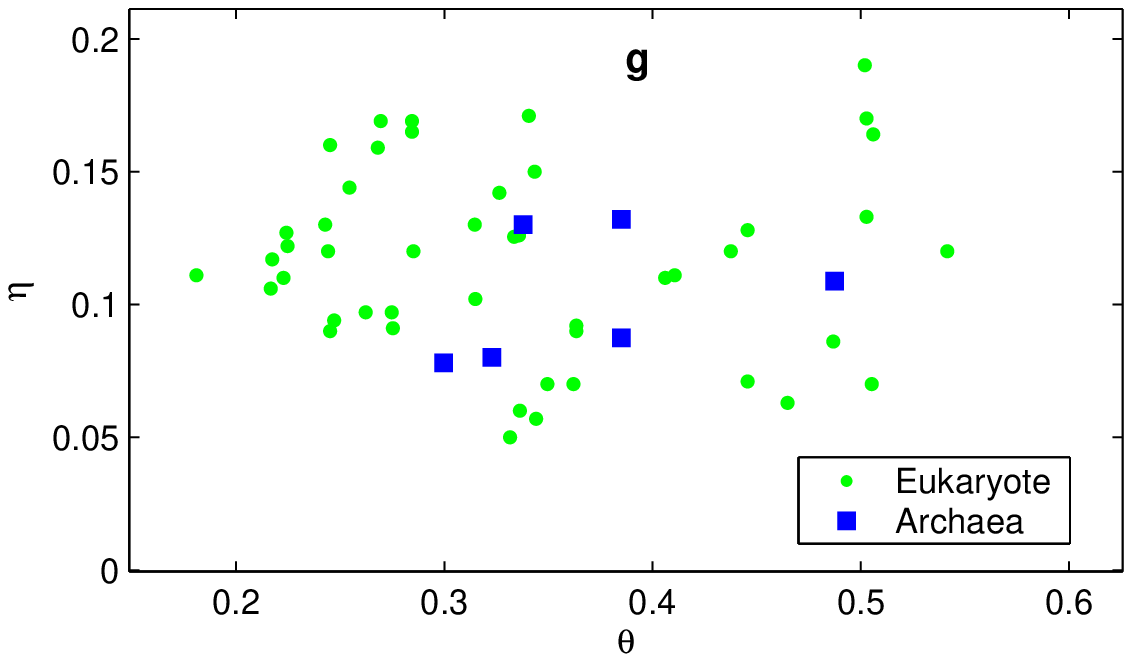}
\includegraphics[width=70mm]{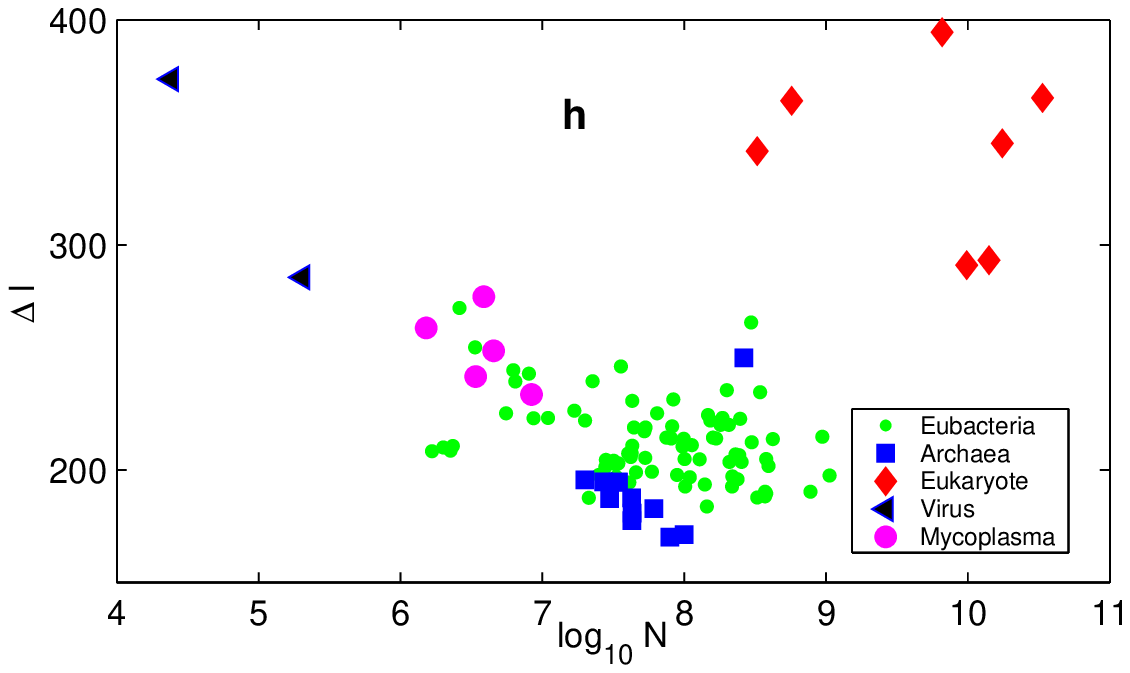}
\includegraphics[width=70mm]{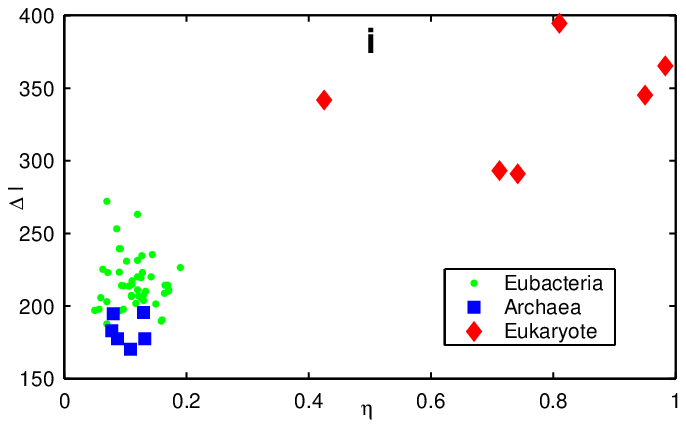}}
\end{figure}

\begin{figure}
\label{fig1} \caption{\small {\bf The mechanism of genome evolution
can be inferred by the circular distribution of species in $\Delta l
-p$ plane.} {\bf a,} The phylogenetic circles consisted of species
in Archaea, Eukarya, Eubacteria and Mycoplasma. The fundamental
relationship goes round in circles in each domain. {\bf b,} Species
only distribute in the left and right quadrants. The origin $O$ is
at $s^* \sim 3.5\times10^6$ bp. {\bf c,} The approximate
proportional relation between $\bar{l}$ and $\Delta l$ (Correlation
coefficient is $0.9022$). {\bf d,} The distribution of species in
$s-\bar{l}$ plane. {\bf e,} The coarse linear relation between
$\bar{l}$ and $\eta$ in each domain. {\bf f,} The distribution of
species in $s-\eta$ plane. {\bf g,} The distribution of species in
$\theta-\eta$ plane. {\bf h,} The distribution of species in
$\log_{10} N -\Delta l$ plane. {\bf h,} Relationship between
non-coding DNA ratio and protein length standard error.}
\end{figure}

\clearpage

\begin{figure}
\centering{
\includegraphics[width=160mm]{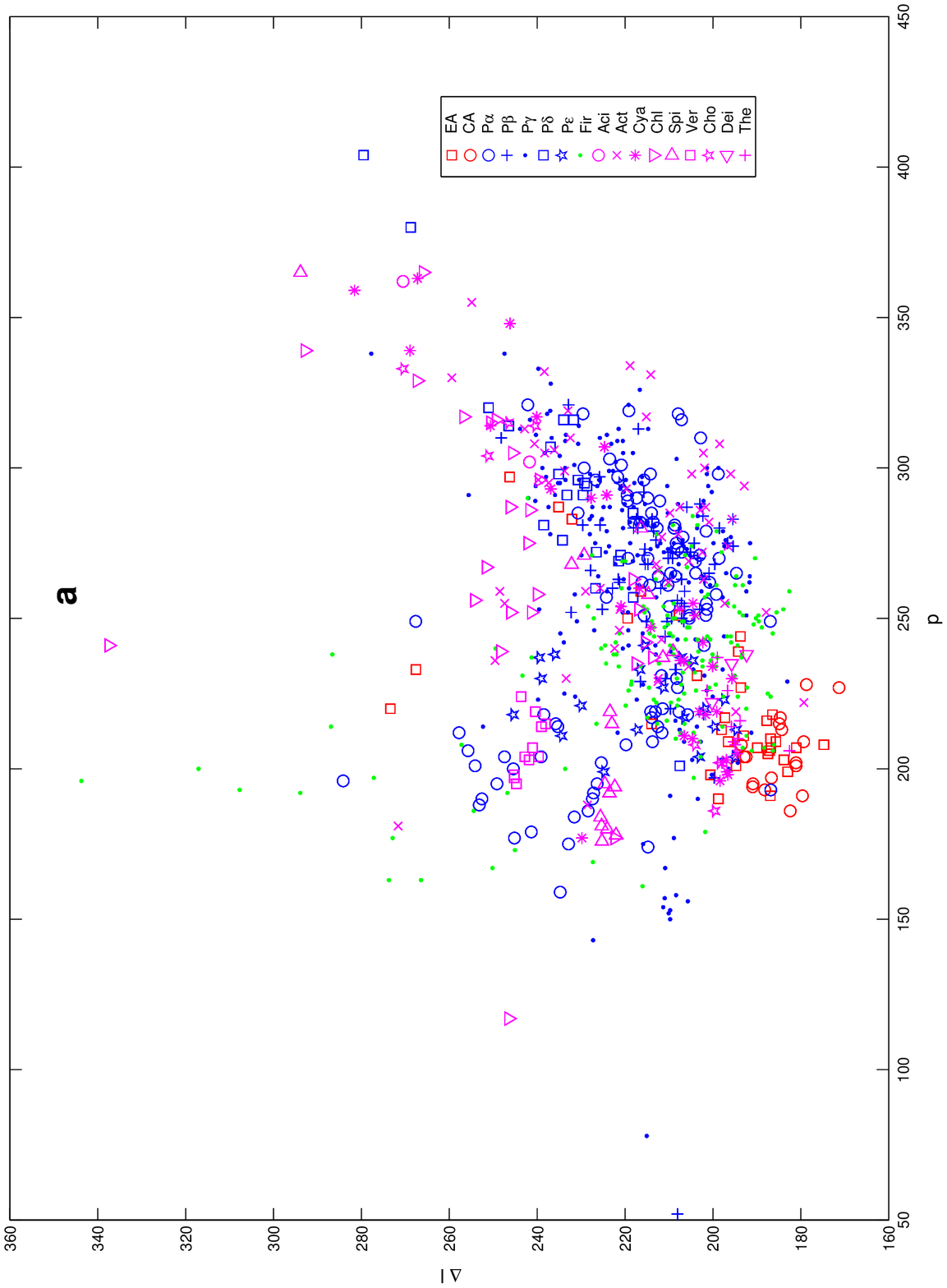}}
\end{figure}

\begin{figure}
\centering{
\includegraphics[width=160mm]{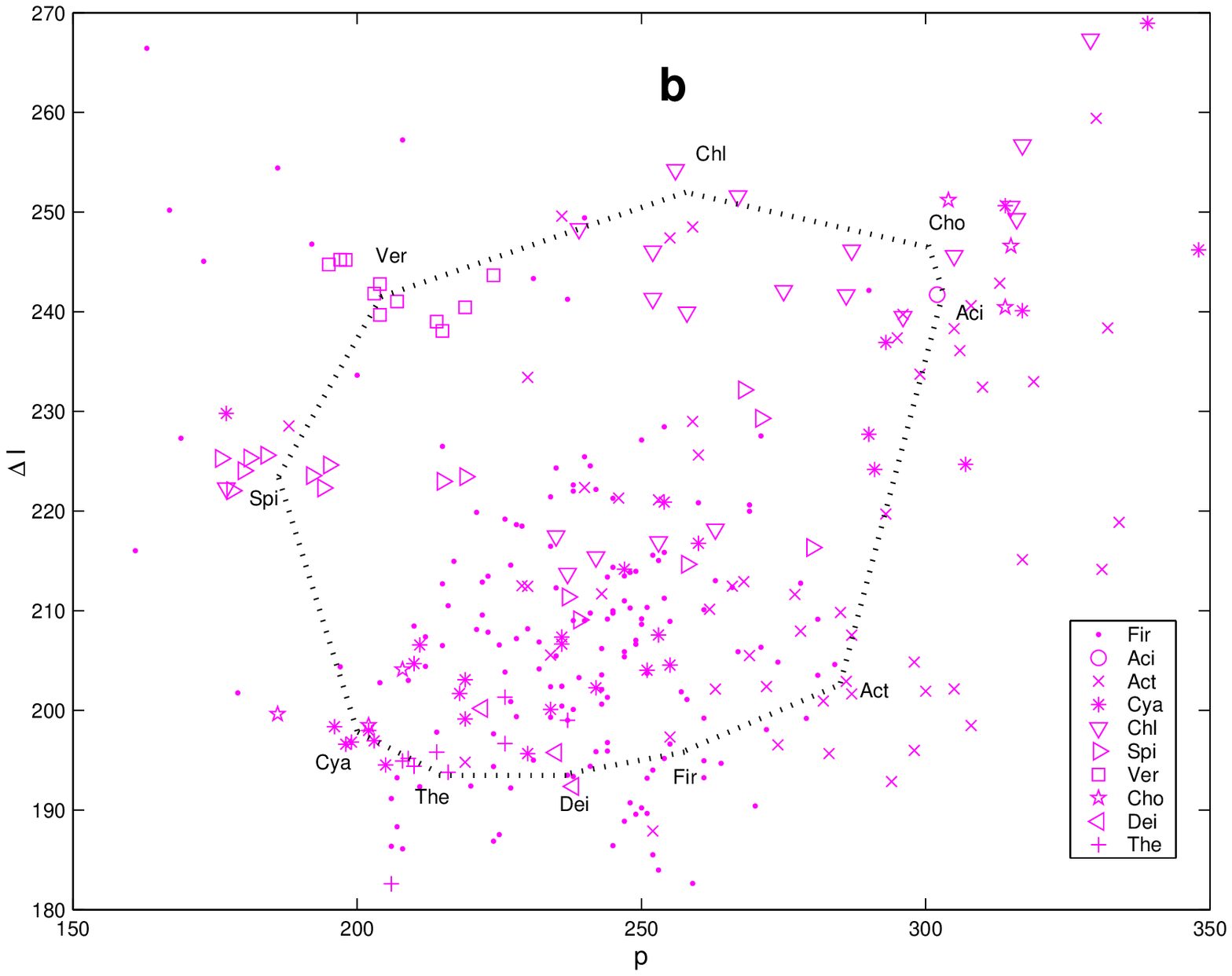}}
\end{figure}

\begin{figure}
\centering{
\includegraphics[width=160mm]{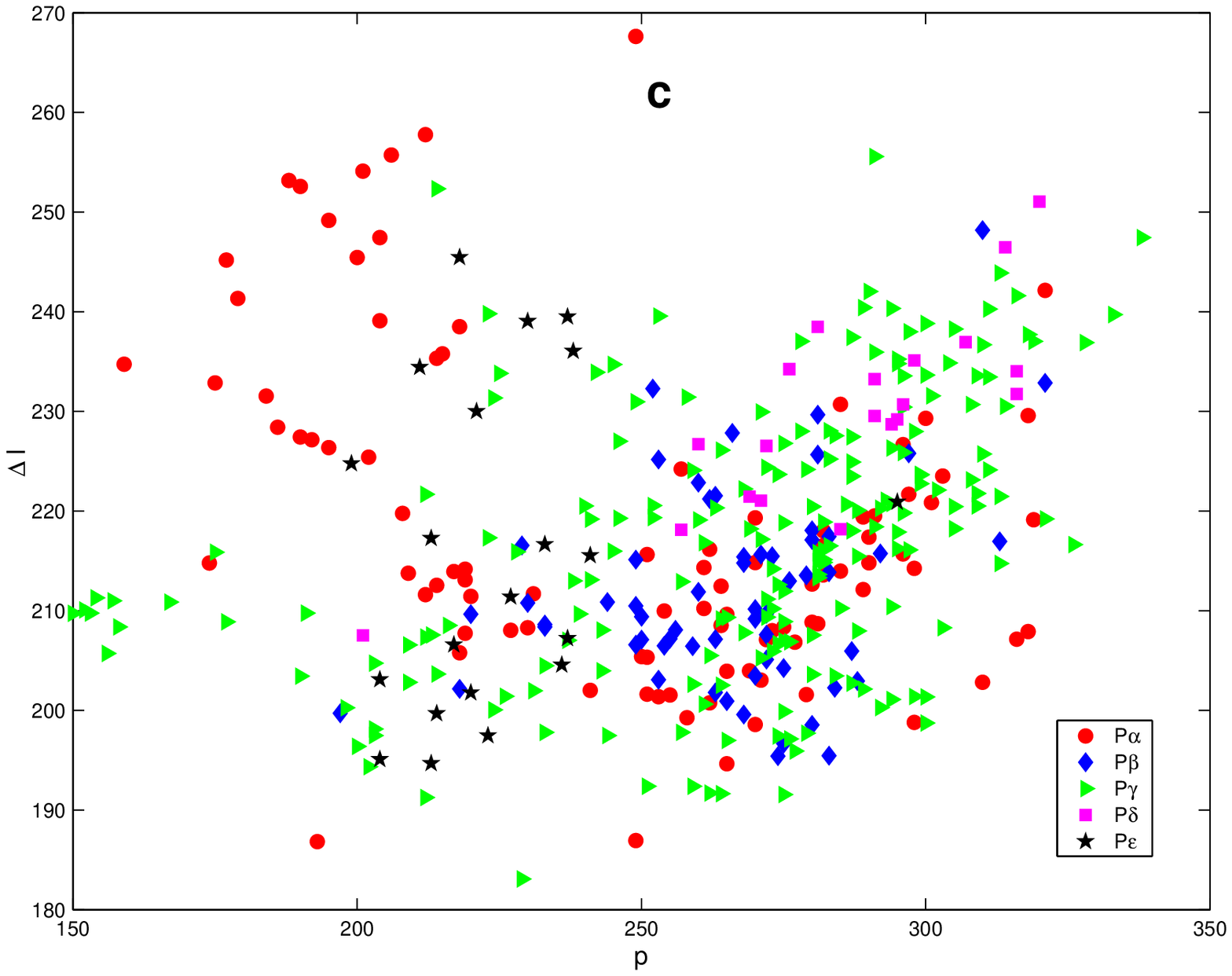}}
\end{figure}

\begin{figure}
\centering{
\includegraphics[width=160mm]{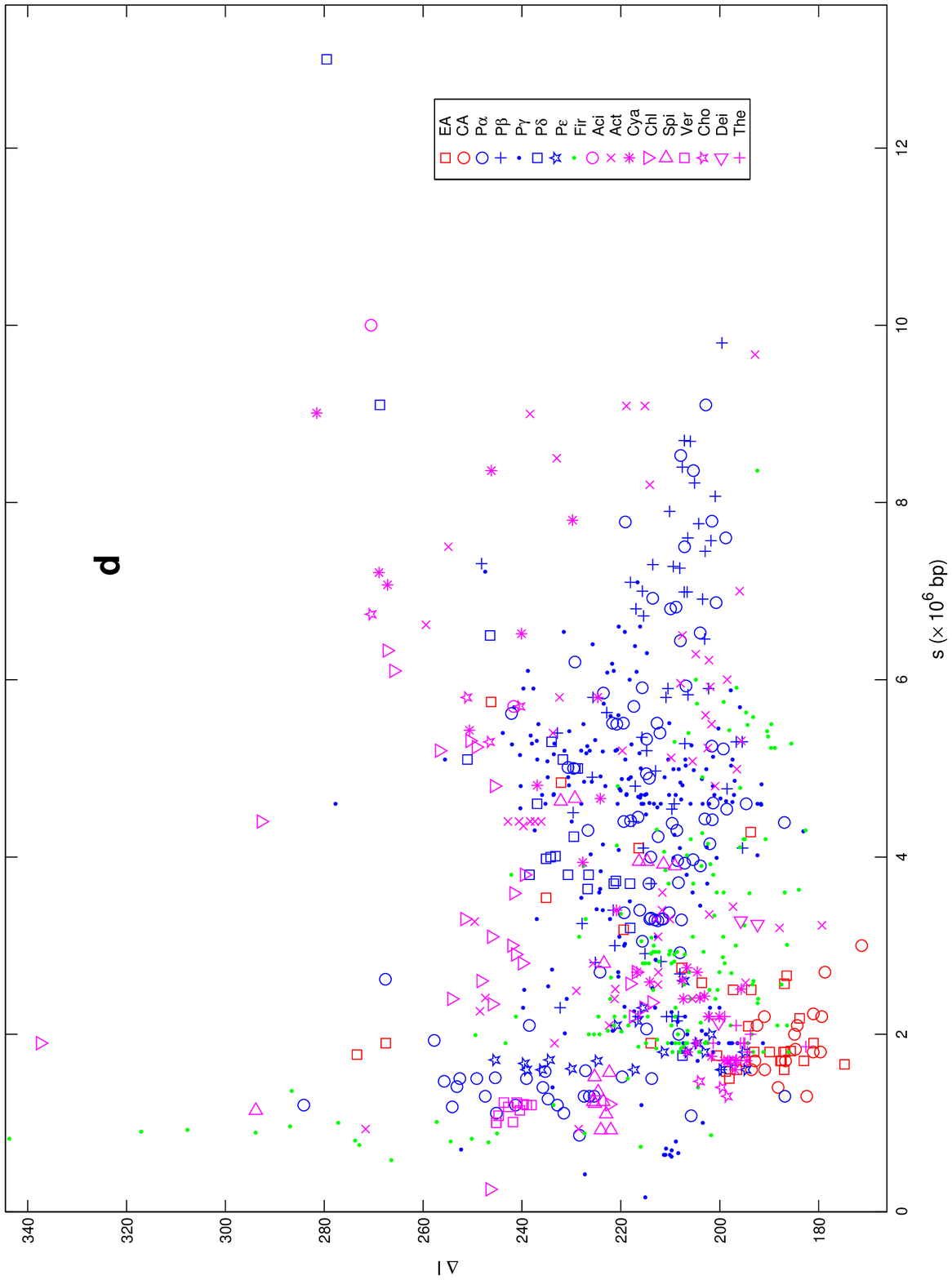}}
\end{figure}

\begin{figure}
\centering{
\includegraphics[width=160mm]{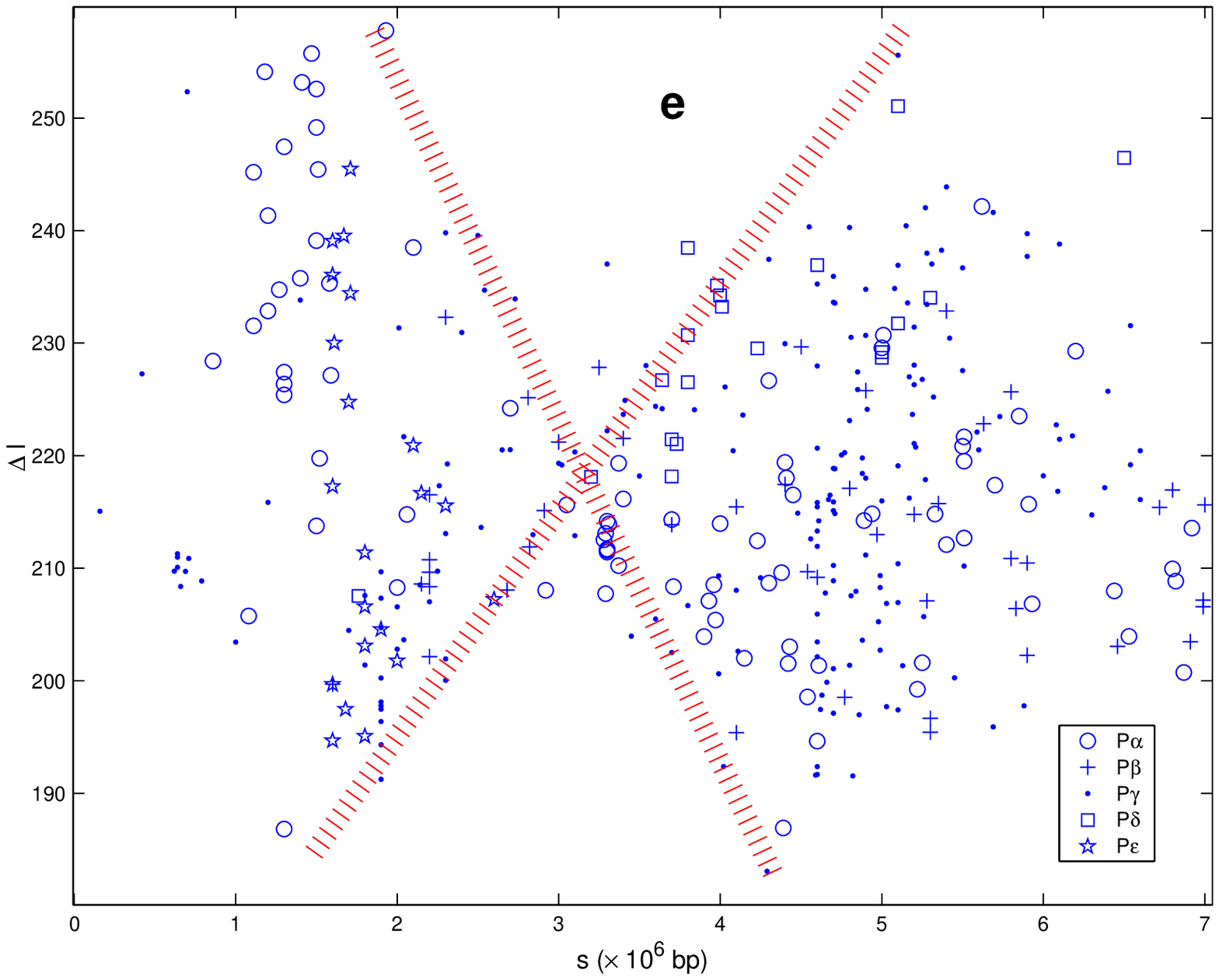}}
\end{figure}

\begin{figure}
\caption{\small {\bf Phylogenetic circles in $\Delta l -p$ plane and
bifurcated distribution in $\Delta l -s$ plane.} The abbreviation of
the names of groups of species are as follows: Euryarchaeota (EA),
Crenarchaeota (CA), Alphaproteobacteria (P$\alpha$),
Betaproteobacteria (P$\beta$), Gammaproteobacteria (P$\gamma$),
Deltaproteobacteria (P$\delta$), Epsilonproteobacteria
(P$\epsilon$), Firmicutes (Fir), Acidobacteria (Aci), Actinobacteria
(Act), Cyanobacteria (Cya), Bacteroidetes/Chlorobi (Chl),
Spirochaetes (Spi), Chlamydiae/Verrucomicrobia (Ver), Chloroflexi
(Cho), Deinococcus-Thermus (Dei), Thermotogae (The). {\bf a,}
Phylogenetic circles formed by $775$ microbes in NCBI. {\bf b,} The
phylogenetic circle formed by phyla of eubacteria. {\bf c,} The
phylogenetic circle formed by proteobacteria. {\bf d,} Bifurcated
distribution of species in $\Delta l -s$ plane. {\bf e,} Bifurcated
distribution of proteobacteria. }
\end{figure}

\clearpage

\begin{figure}
\centering{
\includegraphics[width=160mm]{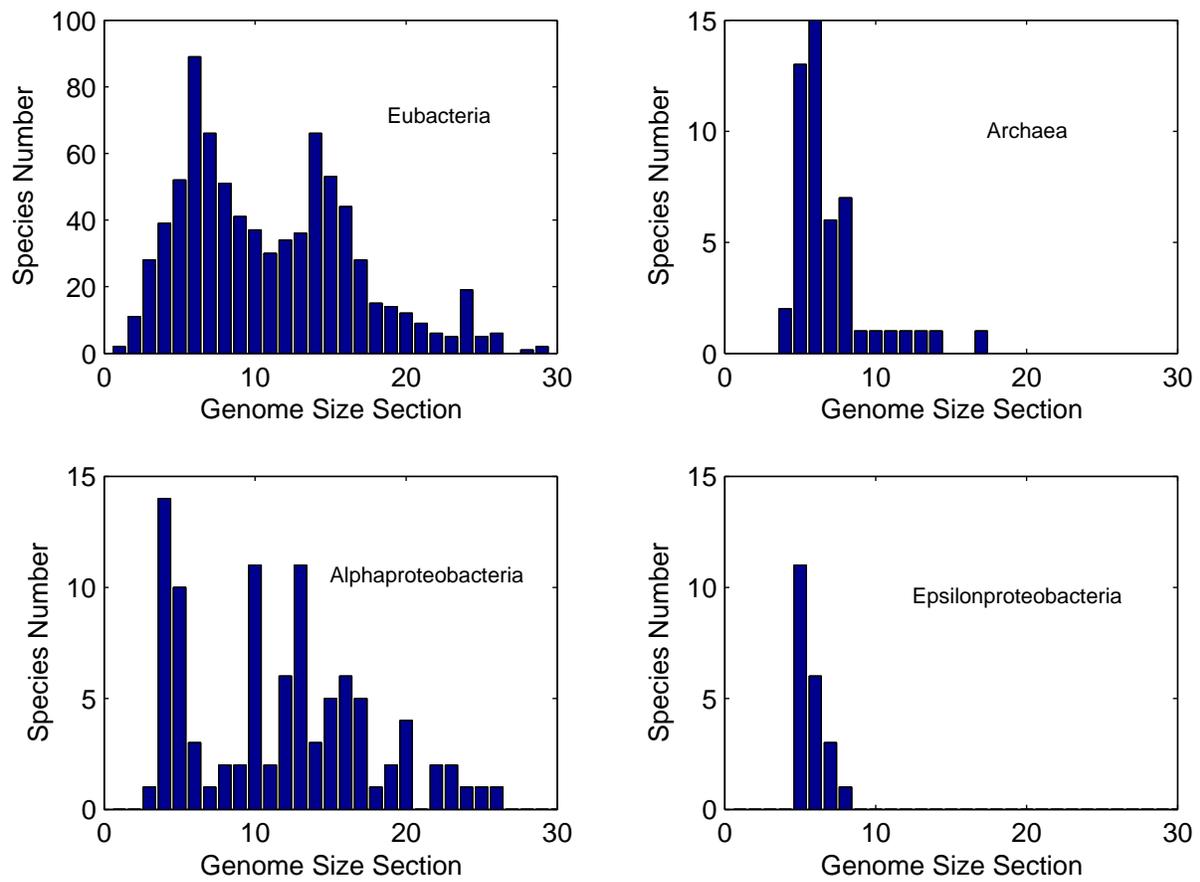}}
\label{fig1} \caption{\small {\bf Genome size distributions.} The
width of each genome size section is $3.5\times10^6$ bp.}
\end{figure}

\clearpage

\begin{figure}
\centering{
\includegraphics[width=70mm]{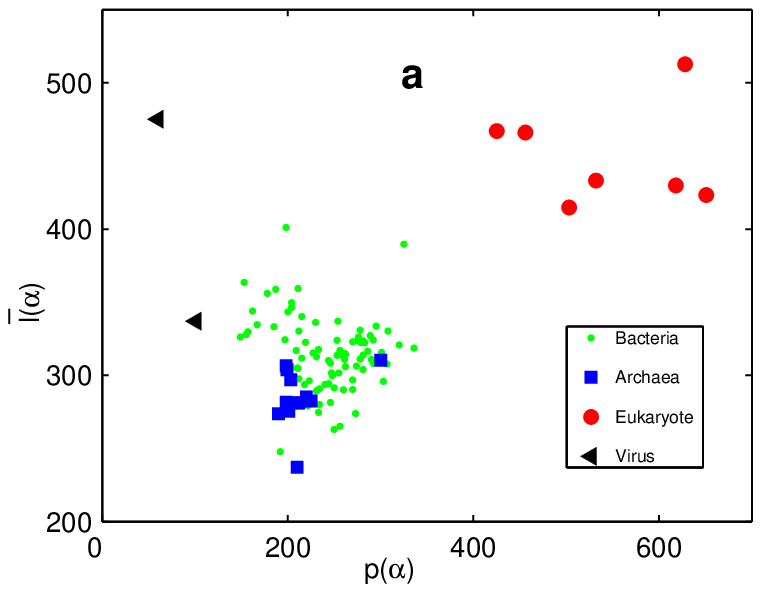}
\includegraphics[width=70mm]{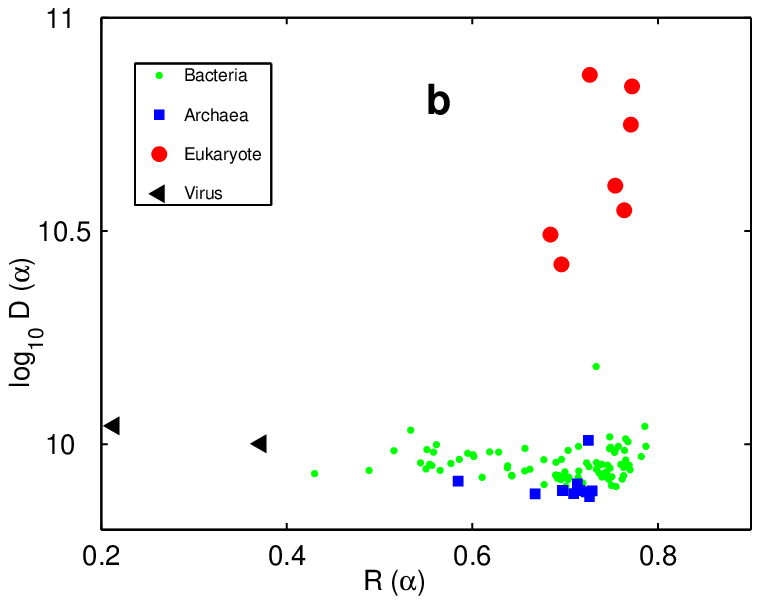}
\includegraphics[width=70mm]{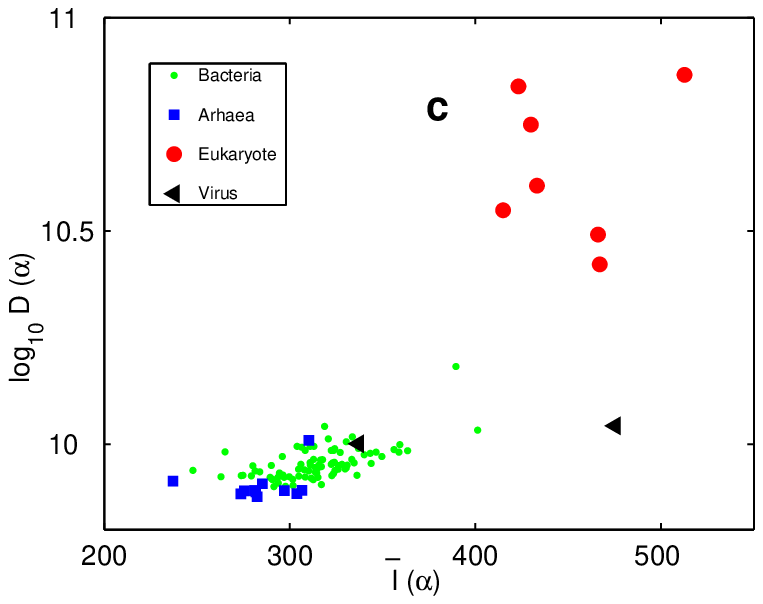}
\includegraphics[width=70mm]{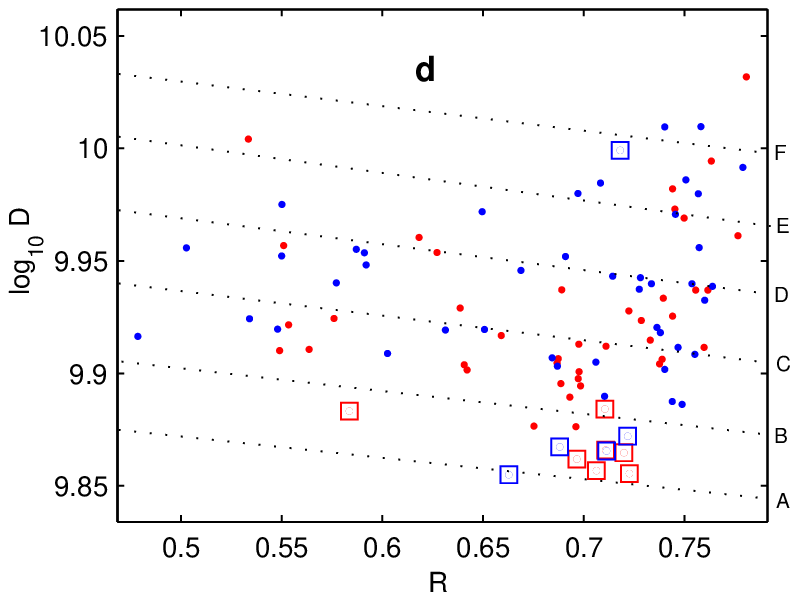}
\includegraphics[width=70mm]{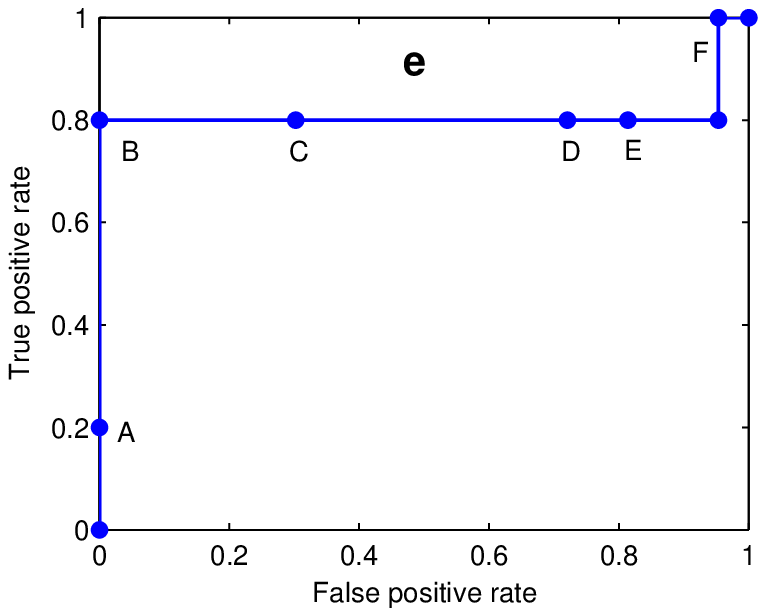}}
\label{fig1} \caption{\small {\bf Classification of life based on
cluster analysis of protein length distributions.} {\bf a,} The
distributions of species in $p-\bar{l}$ plane. {\bf b,} The
distributions of species in $R-\log_{10} D$ plane. {\bf c,} The
distributions of species in $\bar{l}-\log_{10} D$ plane. {\bf d,}
Cross-validation analysis of the classification between Bacteria and
Archaea by the distribtion in $R-\log_{10} D$ plane. The species in
group $G_1$ are red, and the others in $G_2$ are blue. {\bf e,} The
cross-validated ROC curve shows the validity of the classifier.}
\end{figure}

\clearpage
\begin{figure}
\centering{
\includegraphics[width=70mm]{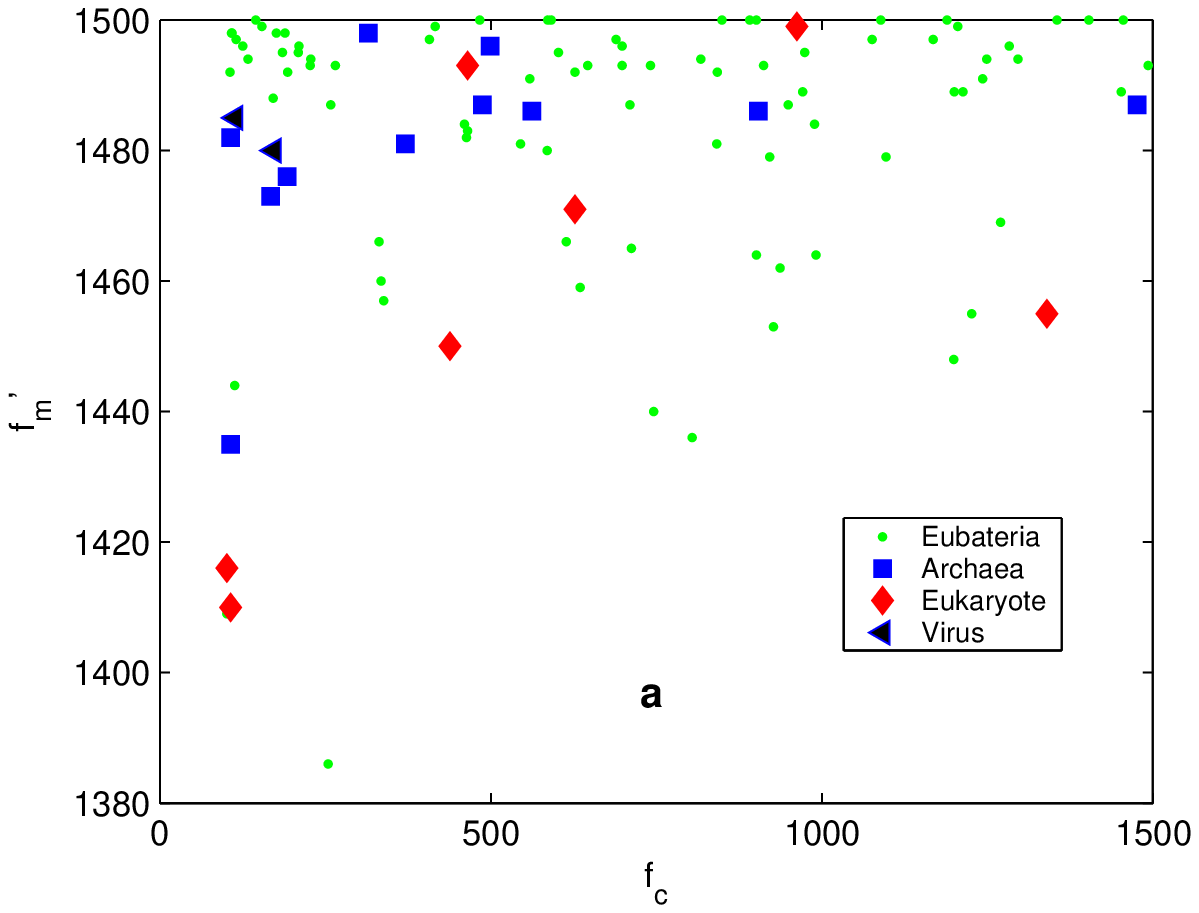}
\includegraphics[width=70mm]{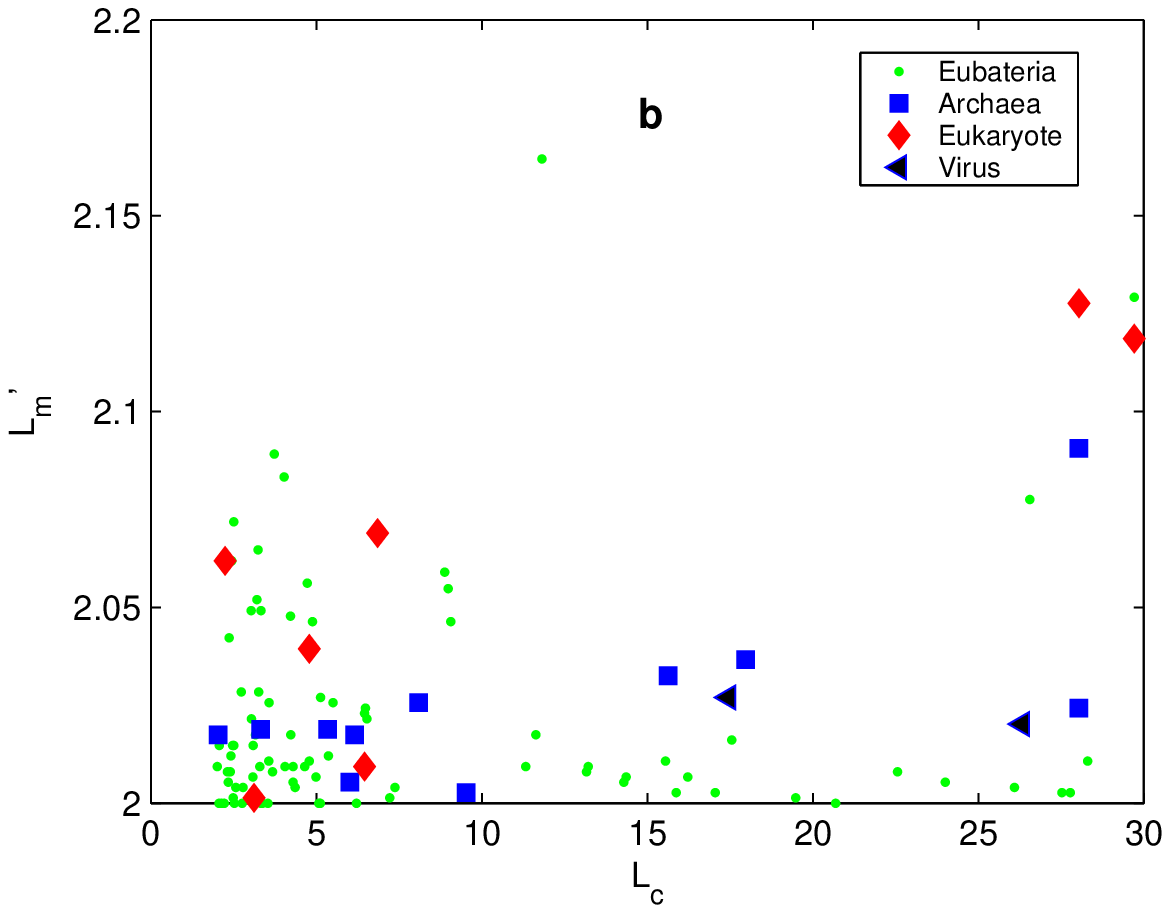}
\includegraphics[width=70mm]{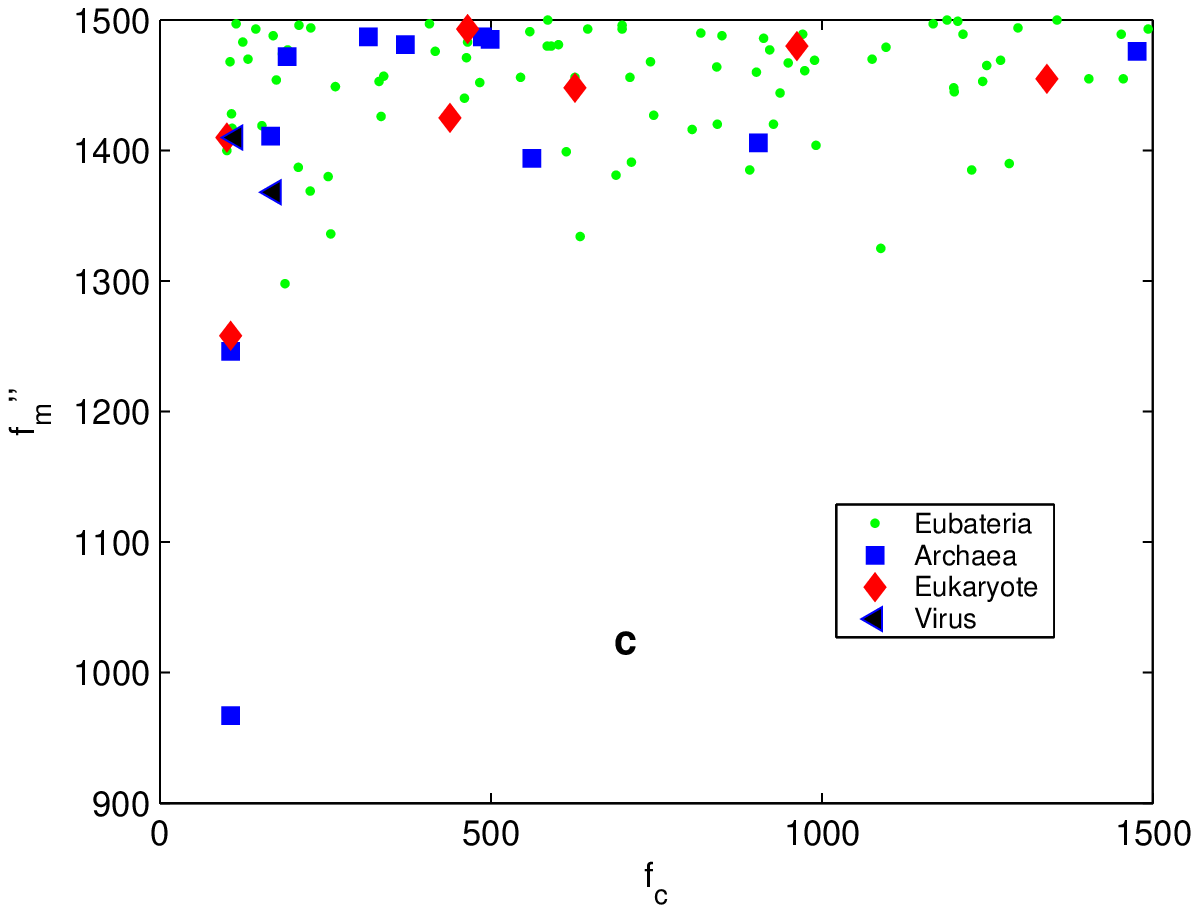}
\includegraphics[width=70mm]{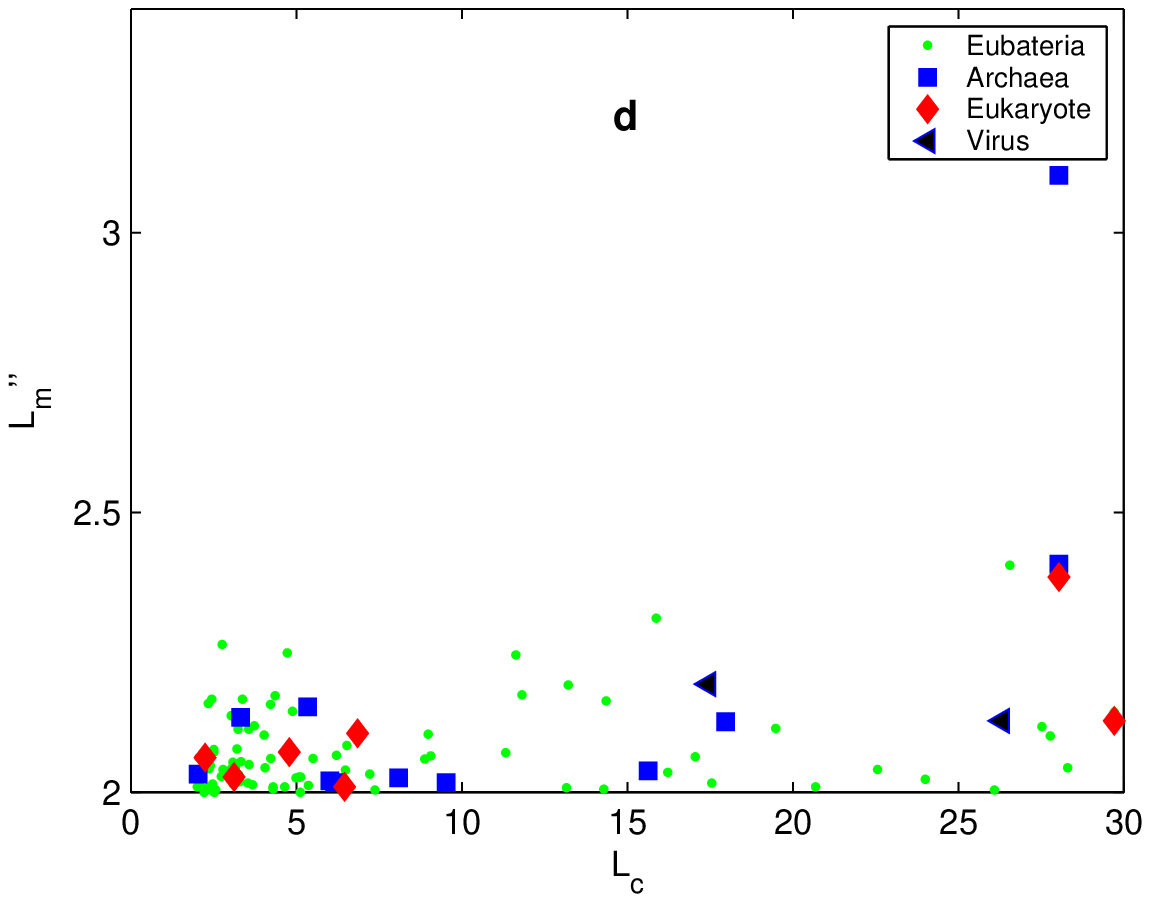}}
\label{fig1} \caption{\small {\bf The protein length hierarchy.} We
choose $n_p=80$ ($n_p=30$) top highest peaks to find the largest
frequency $f_m'$ ($f_m''$) of obvious peaks, and consequently
obtaining $L_m'$ ($L_m''$). {\bf a} $f_m'$ approximately increases
with $f_c$ for Archaea and Eukarya. {\bf b} $L_m'$ approximately
increases with $L_c$ for Archaea and Eukarya. {\bf c} $f_m''$ varies
with $f_c$ in waves for Archaea and Eukarya. {\bf d} $L_m''$ varies
with $L_c$ in waves for Archaea and Eukarya. .}
\end{figure}

\clearpage
\begin{figure}
\centering{
\includegraphics[width=70mm]{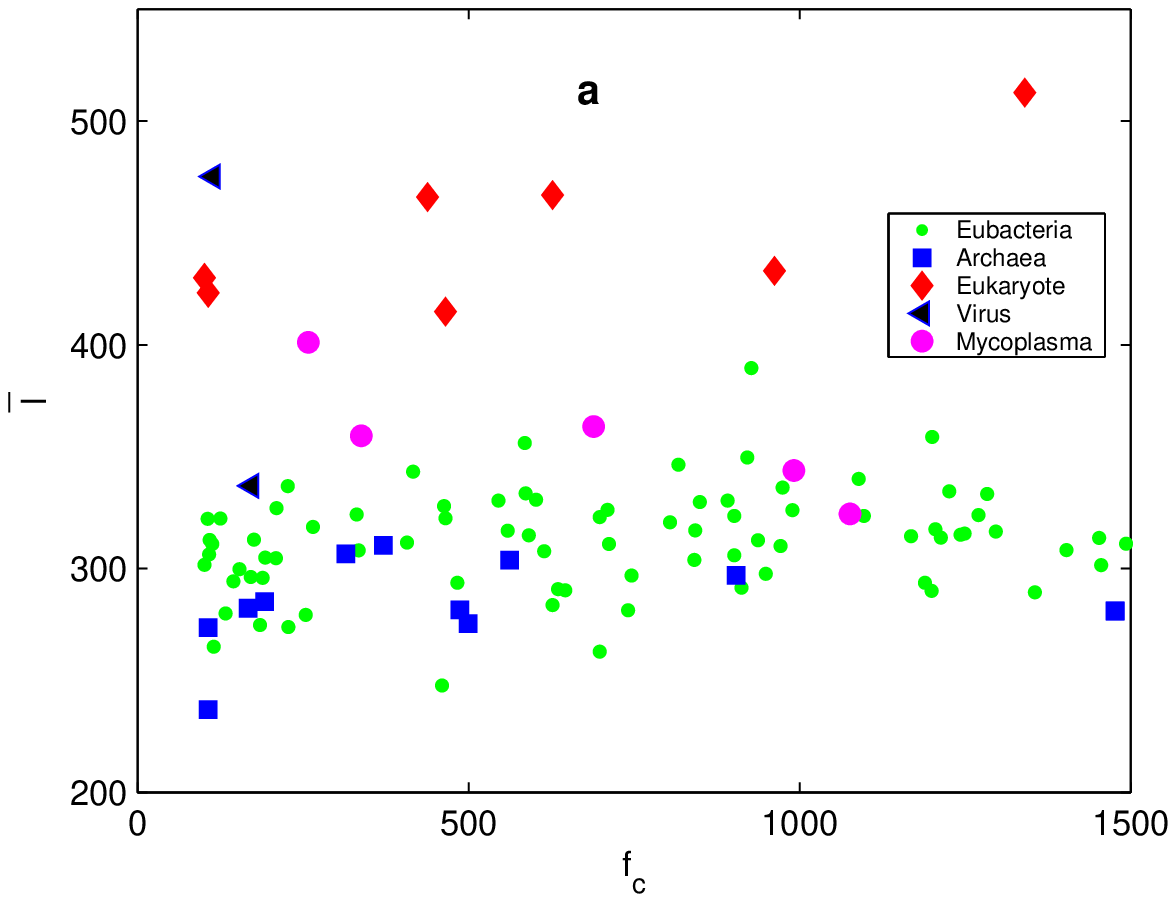}
\includegraphics[width=70mm]{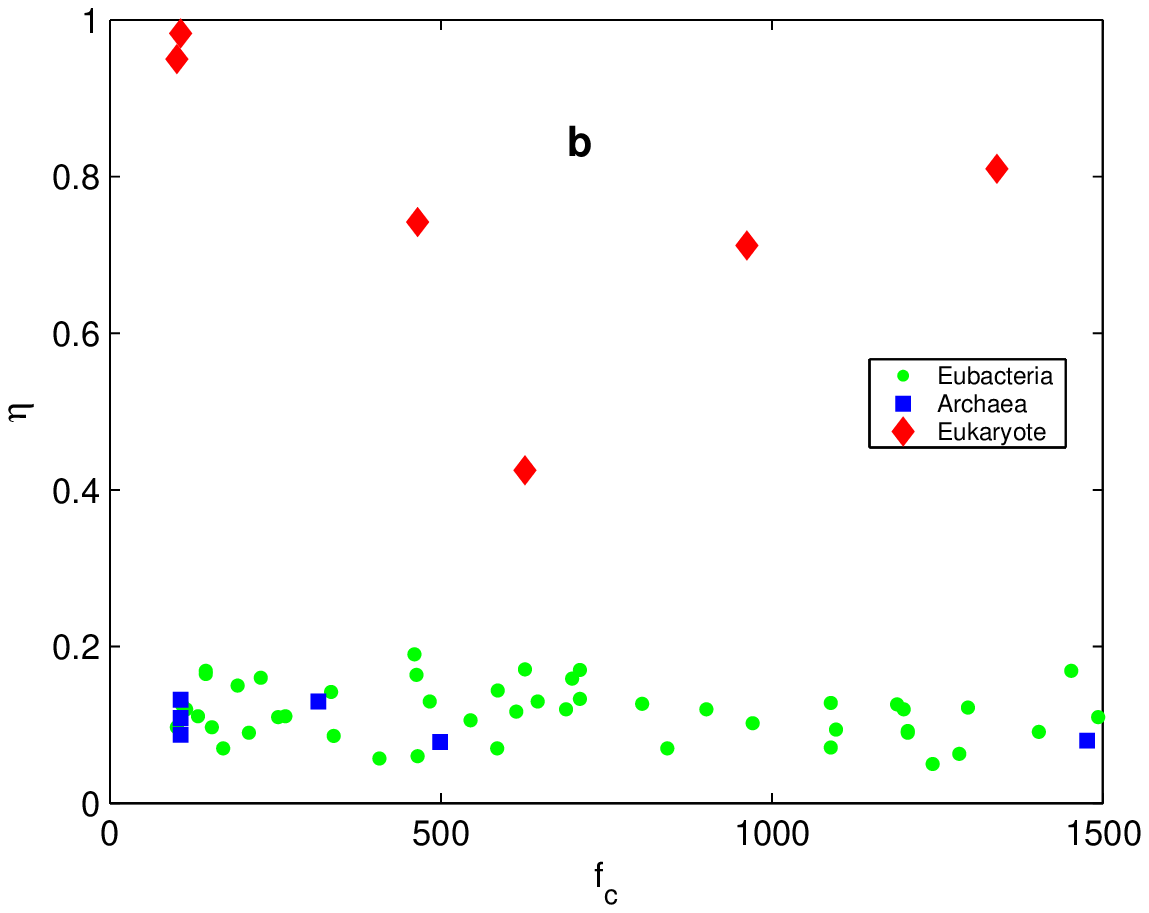}
\includegraphics[width=70mm]{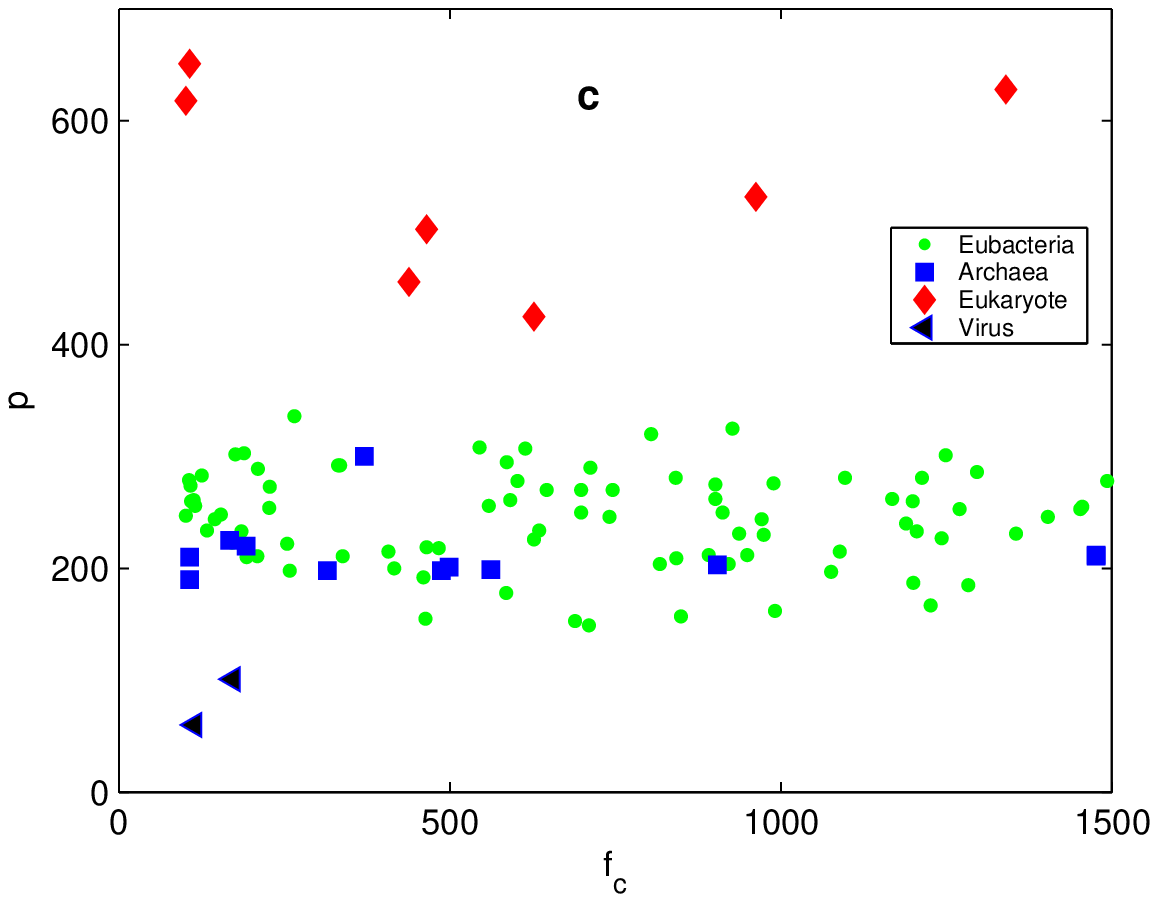}}
\label{fig1} \caption{\small {\bf Explanation of the constraint on
the average protein lengths.} {\bf a,} The rain-bow like
distributions of species in three domains. The arc of Archaea is at
lowest; the arc of Eubacteria is in the middle; and the arc of
Eukarya is on the top. Mycoplasmas also form an arc. {\bf b,} The
relationship between $\eta$ and $f_c$. The V-shaped distribution of
Eukaryotes is obvious. {\bf c,} The relationship between $p$ and
$f_c$. The V-shaped distribution of Eukaryotes is also obvious, and
the distributions of Archaea and Eubacteria form flat arc
respectively.}
\end{figure}

\clearpage

\begin{figure}
\centering{
\includegraphics[width=70mm]{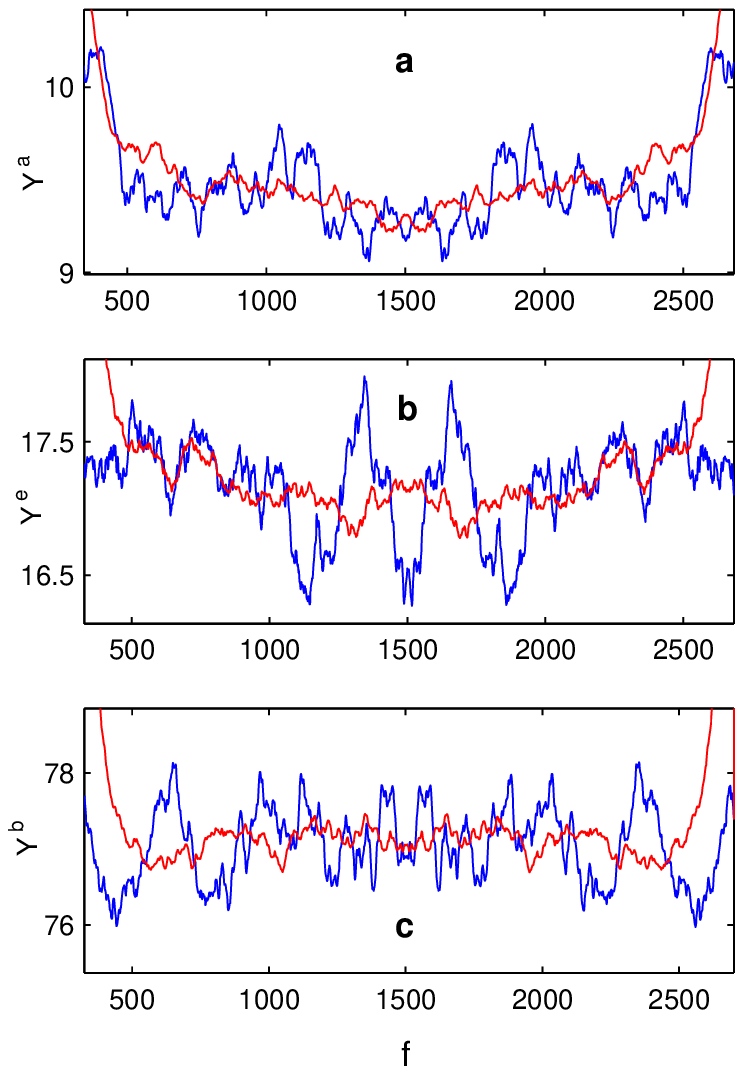}
\includegraphics[width=70mm]{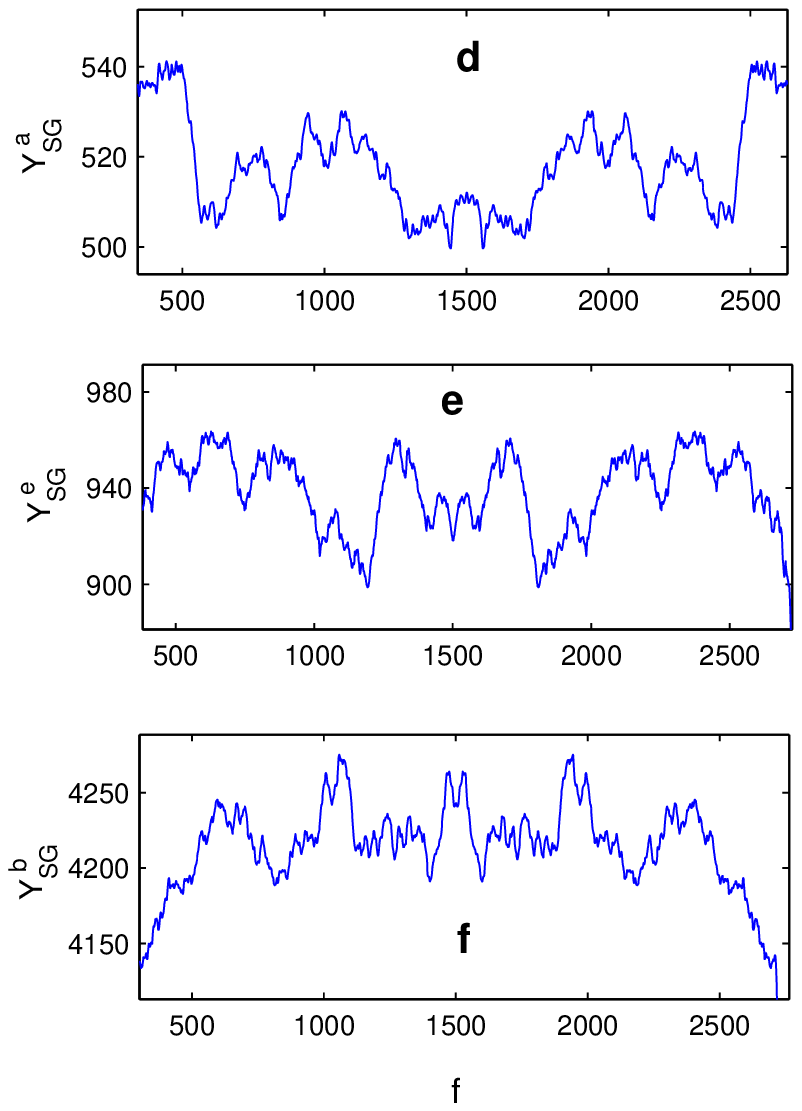}}
\label{fig1} \caption{\small {\bf The profiles of total power
spectra for three domains.} After smoothing, the bottoms of total
power spectra are convex for Archaea and Eucarya but concave for
Bacteria. {\bf a-c,} Smoothing by averaging in neighboring sections.
Neighboring section $w_1=100$ (Blue) and $w_2=300$ (Red). {\bf d-f,}
Smoothing by Savitzky-Golay method (span=501, degree=2).}
\end{figure}

\clearpage

\begin{table}
 \centering
 \caption{The values of properties for species}
\end{table}

 \centering

\begin{tabular}{|r|r|r|r|r|r|r|r|r|r|r|r|}
 \hline
 No. & $\bar{l}$ & $\Delta l$ & $L_c$ & $p$ & $\theta$ & $\eta$ & $S$ & $N$ & $R$ & $D$
 \\ \hline
 1 & 358.83 & 254.48 & 2.50 & 187 & 0.4960 & & & 683 & 0.56 & 9.98 \\
 2 & 314.91 & 204.75 & 5.08 & 261 & 0.2571 & & & 3322 & 0.75 & 9.94 \\
 3 & 237.08 & 170.15 & 28.04 & 210 & 0.4874 & 0.1088 & 1669695 & 2694 & 0.58 & 9.91 \\
 4 & 307.82 & 201.72 & 4.89 & 307 & 0.2173 & 0.1170 & 5674062 & 5402 & 0.77 & 9.94 \\
 5 & 313.81 & 188.23 & 2.47 & 281 & 0.2238 & & & 5274 & 0.76 & 9.95 \\
 6 & 317.02 & 187.64 & 3.56 & 209 & 0.3620 & 0.0700 & 1551335 & 1522 & 0.68 & 9.91 \\
 7 & 433.07 & 293.16 & 3.12 & 532 & 0.2096 & 0.7120 & 115409949 & 25541 & 0.75 & 10.61 \\
 8 & 275.47 & 182.87 & 6.01 & 201 & 0.2996 & 0.0780 & 2178400 & 2406 & 0.73 & 9.89 \\
 9 & 262.96 & 189.58 & 4.30 & 250 & 0.2681 & 0.1590 & 5370060 & 5311 & 0.75 & 9.92 \\
 10 & 273.88 & 190.36 & 13.16 & 273 & 0.2452 & 0.1600 & 546909 & 5274 & 0.76 & 9.93 \\
 11 & 290.35 & 203.62 & 4.64 & 270 & 0.2428 & 0.1300 & 4214810 & 4099 & 0.76 & 9.92 \\
 12 & 389.55 & 265.58 & 3.24 & 325 & 0.2617 & & & 4776 & 0.73 & 10.18 \\
 13 & 304.71 & 221.93 & 14.35 & 211 & 0.3886 & & & 1482 & 0.66 & 9.94 \\
 14 & 330.38 & 223.17 & 3.37 & 212 & 0.4112 & & & 1141 & 0.64 & 9.95 \\
 15 & 324.15 & 221.99 & 9.06 & 292 & 0.2575 & & & 3584 & 0.75 & 9.99 \\
 16 & 322.27 & 187.75 & 28.30 & 279 & 0.2649 & & & 4986 & 0.74 & 9.95 \\
 17 & 333.31 & 225.12 & 2.34 & 185 & 0.4649 & 0.0630 & 1443725 & 850 & 0.59 & 9.96 \\
 18 & 326.10 & 197.07 & 3.03 & 276 & 0.2744 & & & 4184 & 0.73 & 9.94 \\
 19 & 323.53 & 193.58 & 3.33 & 275 & 0.6183 & & & 3446 & 0.43 & 9.93 \\
 20 & 312.96 & 197.49 & 17.05 & 302 & 0.1805 & & & 8307 & 0.79 & 9.99 \\
 21 & 293.74 & 207.45 & 6.21 & 218 & 0.3146 & 0.1300 & 3294935 & 2059 & 0.72 & 9.91 \\
 22 & 328 & 208.60 & 6.48 & 155 & 0.5060 & 0.1640 & 618000 & 574 & 0.55 & 9.95 \\
 23 & 326.21 & 209.97 & 4.23 & 149 & 0.5028 & & & 546 & 0.56 & 9.95 \\
 24 & 329.71 & 208.28 & 3.53 & 157 & 0.5092 & & & 504 & 0.55 & 9.94 \\
 25 & 414.83 & 291.03 & 6.45 & 503 & 0.1945 & 0.7419 & 97000000 & 21832 & 0.76 & 10.55 \\
 26 & 311.59 & 197.79 & 7.37 & 215 & 0.3441 & 0.0570 & 1641181 & 1633 & 0.69 & 9.92 \\
 27 & 334.59 & 210.69 & 2.45 & 167 & 0.5105 & & & 583 & 0.54 & 9.96 \\
 28 & 323.58 & 213.94 & 2.73 & 281 & 0.2471 & 0.0940 & 4016942 & 3737 & 0.75 & 9.99 \\
 29 & 346.45 & 242.84 & 3.67 & 204 & 0.4177 & & & 998 & 0.63 & 9.98 \\
 30 & 343.36 & 239.29 & 7.21 & 200 & 0.4590 & & & 907 & 0.60 & 9.98 \\
 31 & 279.99 & 217.23 & 22.56 & 234 & 0.4107 & 0.1110 & 2154946 & 2252 & 0.64 & 9.95 \\
 32 & 349.61 & 244.35 & 3.26 & 204 & 0.4502 & & & 894 & 0.60 & 9.97 \\
 33 & 311.15 & 206.49 & 2.01 & 278 & 0.2228 & 0.1100 & 4751080 & 4396 & 0.77 & 9.95 \\
 34 & 305.91 & 219.99 & 3.33 & 262 & 0.2442 & 0.1200 & 3940880 & 3847 & 0.76 & 9.95 \\
 35 & 313.70 & 213.96 & 2.07 & 253 & 0.2695 & 0.1690 & 3031430 & 2722 & 0.74 & 9.93 \\
 \hline
\end{tabular}

\begin{tabular}{|r|r|r|r|r|r|r|r|r|r|r|r|}
 \hline
 No. & $\bar{l}$ & $\Delta l$ & $L_c$ & $p$ & $\theta$ & $\eta$ & $S$ & $N$ & $R$ & $D$
 \\ \hline

 36 & 336.16 & 199.14 & 3.08 & 230 & 0.3379 & & & 2373 & 0.69 & 9.93 \\ 37 & 316.99 & 218.83 & 5.37 & 256 & 0.2937 & & & 2269 & 0.73 & 9.95 \\ 38 & 323.04 & 210.43 & 4.30 & 270 & 0.2943 & & & 2947 & 0.72 & 9.96 \\ 39 & 314.44 & 204.83 & 2.57 & 262 & 0.2645 & & & 2989 & 0.74 & 9.93 \\ 40 & 279.28 & 207.24 & 11.81 & 222 & 0.4060 & 0.1100 & 1995275 & 2009 & 0.64 & 9.93 \\ 41 & 308.32 & 196.77 & 2.14 & 246 & 0.2753 & 0.0910 & 3284156 & 3099 & 0.74 & 9.92 \\ 42 & 303.93 & 224.44 & 3.57 & 281 & 0.3107 & & & 3524 & 0.71 & 9.99 \\ 43 & 512.73 & 394.58 & 2.24 & 628 & 0.2562 & 0.8100 & 120000000 & 18358 & 0.73 & 10.87 \\ 44 & 316.53 & 206.93 & 2.31 & 286 & 0.2247 & 0.1220 & 4641000 & 4281 & 0.77 & 9.96 \\ 45 & 290.06 & 211.06 & 2.50 & 260 & 0.2852 & 0.1200 & 3218031 & 3145 & 0.74 & 9.95 \\ 46 & 315.67 & 203.56 & 2.40 & 301 & 0.2226 & & & 4463 & 0.77 & 9.95 \\ 47 & 310.10 & 230.72 & 3.09 & 244 & 0.3149 & 0.1020 & 2714500 & 2067 & 0.71 & 9.94 \\ 48 & 310.96 & 222.74 & 4.21 & 290 & 0.2462 & & & 4425 & 0.76 & 10.00 \\ 49 & 274.78 & 204.43 & 16.22 & 233 & 0.4102 & & & 1715 & 0.64 & 9.93 \\ 50 & 304.90 & 201.23 & 15.54 & 210 & 0.3434 & 0.1500 & 4524893 & 1709 & 0.69 & 9.92 \\ 51 & 285.21 & 187.56 & 15.63 & 220 & 0.3185 & & & 2058 & 0.71 & 9.91 \\ 52 & 336.99 & 285.66 & 17.44 & 101 & 0.6795 & & & 202 & 0.37 & 10.00 \\ 53 & 296.21 & 202.87 & 17.54 & 223 & 0.3495 & 0.0700 & 1799146 & 1874 & 0.69 & 9.93 \\ 54 & 317.57 & 239.38 & 2.49 & 233 & 0.3633 & & & 1564 & 0.68 & 9.96 \\ 55 & 423.24 & 365.33 & 28.04 & 651 & 0.1889 & 0.9830 & 3000000000 & 37229 & 0.77 & 10.84 \\ 56 & 312.62 & 204.01 & 3.20 & 231 & 0.3399 & & & 1813 & 0.70 & 9.92 \\ 57 & 293.62 & 205.33 & 2.52 & 240 & 0.3358 & 0.1260 & 2365589 & 2266 & 0.70 & 9.92 \\ 58 & 301.52 & 192.68 & 2.06 & 255 & 0.2637 & & & 3002 & 0.75 & 9.92 \\ 59 & 297.65 & 194.49 & 3.16 & 212 & 0.3320 & & & 2023 & 0.70 & 9.90 \\ 60 & 310.94 & 214.31 & 26.55 & 261 & 0.2837 & & & 3652 & 0.73 & 9.96 \\ 61 & 299.76 & 213.82 & 19.48 & 248 & 0.2748 & 0.0970 & 3011209 & 2968 & 0.74 & 9.92 \\ 62 & 301.67 & 197.80 & 29.70 & 247 & 0.2622 & 0.0970 & 2944528 & 2833 & 0.75 & 9.90 \\ 63 & 310.32 & 249.91 & 8.09 & 300 & 0.2999 & & & 4540 & 0.72 & 10.01 \\ 64 & 297.05 & 194.72 & 3.32 & 203 & 0.3418 & & & 1687 & 0.70 & 9.89 \\ 65 & 280.99 & 194.59 & 2.03 & 211 & 0.3228 & 0.0800 & 1751377 & 1873 & 0.72 & 9.89 \\ 66 & 281.17 & 194.61 & 2.03 & 212 & 0.3222 & & & 1869 & 0.72 & 9.89 \\ 67 & 429.90 & 345.14 & 29.70 & 618 & 0.1828 & 0.9500 & 2500000000 & 28085 & 0.77 & 10.75 \\ 68 & 475.19 & 373.68 & 26.32 & 60 & 0.8092 & & & 80 & 0.21 & 10.04 \\ 69 & 330.81 & 195.84 & 4.98 & 278 & 0.2537 & & & 4340 & 0.75 & 9.95 \\ 70 & 327.06 & 223.16 & 14.29 & 289 & 0.2451 & 0.0900 & 4345492 & 3906 & 0.75 & 9.98 \\
 \hline
\end{tabular}

\begin{tabular}{|r|r|r|r|r|r|r|r|r|r|r|r|}
 \hline
 No. & $\bar{l}$ & $\Delta l$ & $L_c$ & $p$ & $\theta$ & $\eta$ & $S$ & $N$ & $R$ & $D$
 \\ \hline

 71 & 401.18 & 276.95 & 11.63 & 198 & 0.5086 & & & 726 & 0.53 & 10.03 \\ 72 & 363.49 & 263.10 & 4.35 & 153 & 0.5416 & 0.1200 & 580070 & 484 & 0.52 & 9.98 \\ 73 & 324.39 & 233.57 & 2.79 & 197 & 0.5674 & & & 1016 & 0.49 & 9.94 \\ 74 & 343.90 & 241.56 & 3.03 & 162 & 0.4804 & & & 686 & 0.58 & 9.95 \\ 75 & 359.33 & 253 & 8.88 & 211 & 0.4867 & 0.0860 & 963879 & 778 & 0.56 & 10.00 \\ 76 & 283.77 & 210.78 & 4.78 & 226 & 0.3407 & 0.1710 & 2184406 & 2065 & 0.70 & 9.94 \\ 77 & 323.95 & 225.17 & 2.36 & 253 & 0.3436 & & & 2461 & 0.69 & 9.96 \\ 78 & 291.45 & 183.80 & 3.29 & 250 & 0.2513 & & & 3496 & 0.75 & 9.90 \\ 79 & 336.92 & 245.94 & 13.22 & 254 & 0.3800 & & & 1909 & 0.66 & 9.99 \\ 80 & 330.34 & 213.72 & 5.50 & 308 & 0.2167 & 0.1060 & 6264403 & 5563 & 0.77 & 10.01 \\ 81 & 322.36 & 204.90 & 24 & 283 & 0.2240 & & & 5316 & 0.76 & 9.99 \\ 82 & 303.72 & 187.29 & 5.34 & 199 & 0.3236 & & & 1764 & 0.71 & 9.88 \\ 83 & 281.55 & 180.80 & 6.16 & 198 & 0.3071 & & & 2065 & 0.72 & 9.89 \\ 84 & 273.67 & 177.40 & 28.04 & 190 & 0.3851 & & & 2064 & 0.67 & 9.88 \\ 85 & 320.74 & 234.60 & 3.73 & 320 & 0.2242 & 0.1270 & 5810922 & 5092 & 0.77 & 10.01 \\ 86 & 295.89 & 190.36 & 15.87 & 303 & 0.1953 & & & 7264 & 0.78 & 9.97 \\ 87 & 247.82 & 226.36 & 6.52 & 192 & 0.5019 & 0.1900 & 1268755 & 1374 & 0.57 & 9.94 \\ 88 & 466.99 & 341.69 & 4.78 & 425 & 0.3018 & 0.4250 & 13800000 & 4987 & 0.70 & 10.42 \\ 89 & 296.89 & 192.62 & 4.02 & 270 & 0.4419 & & & 4176 & 0.61 & 9.92 \\ 90 & 294.33 & 214.31 & 20.69 & 244 & 0.2845 & & & 2631 & 0.74 & 9.93 \\ 91 & 289.39 & 198.90 & 2.21 & 231 & 0.3210 & & & 2121 & 0.71 & 9.92 \\ 92 & 318.67 & 214.78 & 11.32 & 336 & 0.1809 & 0.1110 & 8670000 & 7894 & 0.79 & 10.04 \\ 93 & 281.36 & 218.89 & 4.05 & 246 & 0.3949 & & & 2094 & 0.66 & 9.94 \\ 94 & 290.80 & 202.47 & 4.72 & 234 & 0.3350 & & & 1845 & 0.70 & 9.92 \\ 95 & 282.32 & 171.30 & 17.96 & 225 & 0.3006 & & & 2977 & 0.73 & 9.88 \\ 96 & 306.57 & 195.54 & 9.52 & 198 & 0.3378 & 0.1300 & 1564905 & 1478 & 0.70 & 9.89 \\ 97 & 315.18 & 196.90 & 2.41 & 227 & 0.3316 & 0.0500 & 1860725 & 1846 & 0.70 & 9.92 \\ 98 & 340.13 & 222.99 & 2.75 & 215 & 0.4457 & & & 1031 & 0.60 & 9.98 \\ 99 & 356.08 & 272.04 & 5.13 & 178 & 0.5053 & 0.0700 & 751719 & 611 & 0.55 & 9.99 \\ 100 & 312.85 & 219.39 & 27.52 & 260 & 0.3336 & 0.1255 & 4034065 & 2736 & 0.70 & 9.96 \\ 101 & 306.37 & 212.27 & 27.78 & 274 & 0.2561 & & & 4800 & 0.75 & 9.99 \\ 102 & 322.57 & 205.70 & 6.45 & 219 & 0.3362 & 0.0600 & 2110355 & 2044 & 0.69 & 9.93 \\ 103 & 333.68 & 235.42 & 5.12 & 295 & 0.2545 & 0.1440 & 5175554 & 4029 & 0.75 & 10.02 \\ 104 & 265.15 & 231.31 & 26.09 & 256 & 0.4376 & 0.1200 & 2679305 & 2763 & 0.62 & 9.98 \\ 105 & 466.08 & 364.04 & 6.85 & 456 & 0.3221 & & & 6356 & 0.68 & 10.49 \\ 106 & 308.20 & 220.05 & 8.98 & 292 & 0.3265 & 0.1420 & 4653728 & 4087 & 0.70 & 9.99 \\

 \hline
\end{tabular}

\clearpage

\begin{table}
 \centering
 \caption{List of the species in PEP}
\end{table}

\centering
\begin{tabular}{|l|}
 \hline (No. 1) {\tiny } Acholeplasma florum (Mesoplasma florum)
{\tiny DOMAIN:} Eubacteria \\ \hline (No. 2) Acinetobacter sp
(strain ADP1) {\tiny DOMAIN:} Eubacteria
\\ \hline
(No. 3) Aeropyrum pernix K1 {\tiny DOMAIN:} Archaebacteria \\
\hline (No. 4) Agrobacterium tumefaciens (strain C58 / ATCC 33970)
{\tiny Eubacteria} \\ \hline (No. 5)
 Agrobacterium tumefaciens {\tiny DOMAIN:} Eubacteria \\
\hline (No. 6) Aquifex aeolicus {\tiny DOMAIN:} Eubacteria \\
\hline (No. 7) Arabidopsis thaliana {\tiny DOMAIN:} Eukaryote
\\ \hline (No.
8) Achaeoglobus fulgidus {\tiny DOMAIN:} Archaebacteria \\
\hline (No. 9) Bacillus anthracis (strain Ames) {\tiny DOMAIN:}
Eubacteria \\ \hline (No. 10)
Bacillus cereus (ATCC 14579) {\tiny DOMAIN:} Eubacteria \\
\hline
(No. 11) Bacillus subtilis {\tiny DOMAIN:} Eubacteria \\
\hline (No. 12) Bacteroides thetaiotaomicron VPI-5482 {\tiny
DOMAIN:} Eubacteria \\ \hline (No. 13) Bartonella henselae
(Houston-1) {\tiny DOMAIN:} Eubacteria
\\ \hline (No. 14) Bartonella quintana (Toulouse)
{\tiny DOMAIN:} Eubacteria
 \\ \hline (No. 15) Bdellovibrio bacteriovorus {\tiny DOMAIN:}
Eubacteria \\ \hline (No. 16) Bordetella bronchiseptica RB50 {\tiny
DOMAIN:} Eubacteria \\ \hline (No. 17) Borrelia burgdorferi {\tiny
DOMAIN:} Eubacteria
\\ \hline (No.
18) Bordetella parapertussis {\tiny DOMAIN:} Eubacteria \\
\hline
(No. 19) Bordetella pertussis {\tiny DOMAIN:} Eubacteria \\
\hline (No. 20) Bradyrhizobium japonicum {\tiny DOMAIN:} Eubacteria
 \\ \hline
(No. 21) Brucella melitensis; B melitensis; brume {\tiny DOMAIN:}
Eubacteria \\ \hline (No. 22) Buchnera aphidicola (subsp.
Acyrthosiphon pisum) {\tiny Eubacteria}
 \\ \hline (No. 23) Buchnera aphidicola (subsp. Schizaphis
graminum) {\tiny Eubacteria} \\ \hline (No. 24) Buchnera aphidicola
(subsp. Baizongia pistaciae) {\tiny Eubacteria}
\\ \hline (No. 25) Caenorhabditis elegans {\tiny DOMAIN:}
Eukaryote \\ \hline (No. 26) Campylobacter jejuni {\tiny DOMAIN:}
Eubacteria \\ \hline (No. 27) Candidatus Blochmannia floridanus
{\tiny DOMAIN:} Eubacteria
\\ \hline
(No. 28) Caulobacter crescentus {\tiny DOMAIN:} Eubacteria \\
\hline (No. 29) Chlamydophila caviae {\tiny DOMAIN:} Eubacteria
 \\ \hline (No. 30) Chlamydia muridarum {\tiny DOMAIN:} Eubacteria
 \\ \hline
(No. 31) Chlorobium tepidum {\tiny DOMAIN:} Eubacteria
 \\ \hline (No. 32) Chlamydia trachomatis {\tiny DOMAIN:} Eubacteria
 \\ \hline (No. 33) Chromobacterium violaceum ATCC 12472
{\tiny DOMAIN:} Eubacteria \\ \hline (No. 34) Clostridium
acetobutylicum {\tiny DOMAIN:} Eubacteria \\ \hline (No.
35) Clostridium perfringens {\tiny DOMAIN:} Eubacteria \\
\hline

\end{tabular}
\clearpage
\begin{tabular}{|l|}
\hline

 (No. 36) Clostridium tetani {\tiny DOMAIN:}
Eubacteria \\ \hline (No. 37)
Corynebacterium diphtheriae NCTC 13129\ {\tiny DOMAIN:} Eubacteria \\
\hline (No. 38)
 Corynebacterium efficiens {\tiny DOMAIN:} Eubacteria \\
\hline (No. 39) Corynebacterium glutamicum {\tiny DOMAIN:}
Eubacteria \\ \hline (No. 40) Coxiella burnetii {\tiny DOMAIN:}
Eubacteria \\ \hline (No. 41) Deinococcus radiodurans {\tiny
DOMAIN:} Eubacteria
 \\ \hline (No. 42) Desulfovibrio vulgaris subsp. vulgaris
str. Hildenborough {\tiny Eubacteria} \\ \hline (No. 43)
 Drosophila melanogaster {\tiny DOMAIN:} Eukaryote
\\ \hline
(No. 44) Escherichia coli {\tiny DOMAIN:} Eubacteria \\
\hline (No. 45) Enterococcus faecalis {\tiny DOMAIN:} Eubacteria
\\ \hline (No. 46) Erwinia carotovora {\tiny DOMAIN:} Eubacteria
 \\ \hline (No. 47) Fusobacterium nucleatum {\tiny DOMAIN:}
Eubacteria \\ \hline (No. 48) Gloeobacter violaceus {\tiny DOMAIN:}
Eubacteria \\ \hline (No. 49) Haemophilus ducreyi {\tiny DOMAIN:}
Eubacteria \\ \hline (No.
50) Haemophilus influenzae {\tiny DOMAIN:} Eubacteria \\
\hline (No. 51) Halobacterium sp. (strain NRC-1) {\tiny DOMAIN:}
Archaebacteria \\ \hline (No. 52)
Human cytomegalovirus (strain AD169) {\tiny DOMAIN:} virus \\
\hline (No. 53)
 Helicobacter heilmannii {\tiny DOMAIN:} Eubacteria \\
\hline
(No. 54) Helicobacter pylori {\tiny DOMAIN:} Eubacteria \\
\hline (No. 55) Homo sapiens {\tiny DOMAIN:} Eukaryote \\
\hline (No. 56) Lactobacillus johnsonii {\tiny DOMAIN:} Eubacteria
 \\ \hline (No. 57) Lactococcus lactis (subsp. lactis)
{\tiny DOMAIN:} Eubacteria \\ \hline (No. 58)
Lactobacillus plantarum WCFS1 {\tiny DOMAIN:} Eubacteria \\
\hline (No. 59) Leifsonia xyli (subsp. xyli) {\tiny DOMAIN:}
Eubacteria
\\ \hline (No. 60) {\tiny Leptospira interrogans (serogroup
Icterohaemorrhagiae / serovar Copenhageni)} {\tiny DOMAIN:}
Eubacteria \\ \hline
(No. 61) Listeria innocua {\tiny DOMAIN:} Eubacteria \\
\hline (No. 62) Listeria monocytogenes {\tiny DOMAIN:} Eubacteria
\\ \hline (No. 63) Methanosarcina acetivorans
{\tiny DOMAIN:} Archaebacteria \\ \hline (No. 64) Methanopyrus kandleri {\tiny DOMAIN:} Archaebacteria \\
\hline (No. 65) Methanobacterium thermoautotrophicum {\tiny DOMAIN:}
Archaebacteria \\ \hline (No. 66)
Methanobacterium thermoautotrophicum {\tiny DOMAIN:} Archaebacteria \\
\hline (No. 67) Mus musculus {\tiny DOMAIN:} Eukaryote \\
\hline (No. 68) Murine herpesvirus 68 strain WUMS {\tiny DOMAIN:}
virus
 \\ \hline (No. 69) Mycobacterium avium; M avium; mycav
{\tiny DOMAIN:} Eubacteria \\ \hline (No. 70) Mycobacterium bovis
AF2122/97 {\tiny DOMAIN:} Eubacteria \\ \hline

\end{tabular}
\clearpage
\begin{tabular}{|l|}
\hline

(No. 71) Mycoplasma gallisepticum {\tiny DOMAIN:} Eubacteria \\
\hline (No. 72) Mycoplasma genitalium {\tiny DOMAIN:} Eubacteria
 \\ \hline (No. 73) Mycoplasma mycoides (subsp. mycoides SC)
{\tiny DOMAIN:} Eubacteria \\ \hline (No. 74) Mycoplasma pneumoniae
{\tiny DOMAIN:} Eubacteria \\ \hline (No. 75)
 Mycoplasma pulmonis {\tiny DOMAIN:} Eubacteria \\
\hline (No. 76) Neisseria meningitidis {\tiny DOMAIN:} Eubacteria \\
\hline (No. 77) Nitrosomonas europaea {\tiny DOMAIN:} Eubacteria
 \\ \hline (No. 78) Oceanobacillus iheyensis {\tiny DOMAIN:}
Eubacteria \\ \hline (No. 79) Porphyromonas gingivalis {\tiny
DOMAIN:} Eubacteria \\ \hline (No. 80) Pseudomonas aeruginosa {\tiny DOMAIN:} Eubacteria \\
\hline (No. 81) Pseudomonas putida {\tiny DOMAIN:} Eubacteria
 \\ \hline (No. 82) Pyrococcus abyssi {\tiny DOMAIN:} Archaebacteria
 \\ \hline (No. 83) Pyrococcus furiosus {\tiny DOMAIN:}
Archaebacteria \\ \hline (No. 84) Pyrococcus horikoshii {\tiny
DOMAIN:} Archaebacteria \\ \hline (No. 85)
 Ralstonia solanacearum {\tiny DOMAIN:} Eubacteria \\
\hline
(No. 86) Rhizobium loti {\tiny DOMAIN:} Eubacteria \\
\hline
(No. 87) Rickettsia conorii {\tiny DOMAIN:} Eubacteria \\
\hline (No. 88) Schizosaccharomyces pombe {\tiny DOMAIN:} Eukaryote
\\ \hline (No. 89) Shigella flexneri Shigella
flexneri {\tiny DOMAIN:} Eubacteria \\ \hline (No. 90) Staphylococcus aureus {\tiny DOMAIN:} Eubacteria \\
\hline (No. 91) Streptococcus agalactiae {\tiny DOMAIN:} Eubacteria
 \\ \hline (No. 92) Streptomyces coelicolor {\tiny DOMAIN:}
Eubacteria \\ \hline (No. 93) Streptococcus pneumoniae {\tiny
DOMAIN:} Eubacteria \\ \hline (No. 94) Streptococcus pyogenes {\tiny DOMAIN:} Eubacteria \\
\hline (No. 95)
 Sulfolobus solfataricus {\tiny DOMAIN:} Archaebacteria \\
\hline (No. 96) Thermoplasma acidophilum {\tiny DOMAIN:}
Archaebacteria
 \\ \hline (No. 97) Thermotoga maritima {\tiny DOMAIN:}
Eubacteria \\ \hline (No. 98) Treponema pallidum {\tiny DOMAIN:}
Eubacteria \\ \hline (No. 99) Ureaplasma urealyticum {\tiny DOMAIN:} Eubacteria \\
\hline (No. 100) Vibrio cholerae {\tiny DOMAIN:} Eubacteria
\\ \hline
(No. 101) Vibrio parahaemolyticus RIMD 2210633 {\tiny DOMAIN:}
Eubacteria \\ \hline (No. 102) Wolinella succinogenes {\tiny
DOMAIN:} Eubacteria \\ \hline (No. 103) {\tiny Xanthomonas
axonopodis (pv. citri); X axonopodis (pv. citri)} {\tiny DOMAIN:}
Eubacteria \\ \hline (No. 104) Xylella fastidiosa {\tiny DOMAIN:}
Eubacteria \\ \hline (No. 105) Saccharomyces cerevisiae {\tiny
DOMAIN:} Eukaryote \\ \hline (No. 106) Yersinia pestis {\tiny
DOMAIN:} Eubacteria
\\ \hline

\end{tabular}

\end{document}